\documentclass[twocolumn,showpacs,showkeys,preprintnumbers,amsmath,amssymb,superscriptaddress]{revtex4}

\usepackage{graphicx}
\usepackage{dcolumn}
\usepackage{bm}

\begin{document}
\topmargin=-5 true mm

\preprint{}

\title{Emergence of spatiotemporal chaos driven by far-field breakup of spiral waves in the plankton ecological systems}
\author{Quan-Xing Liu}
\affiliation {%
Department of Mathematics, North University of China, Taiyuan,
Shan'xi 030051, People's Republic of China
}%
\author{Gui-Quan Sun}
\affiliation {%
Department of Mathematics, North University of China, Taiyuan,
Shan'xi 030051, People's Republic of China
}%
\author{Bai-Lian Li}%
\affiliation{Ecological Complexity and Modeling Laboratory,
Department of Botany and Plant Sciences,
University of California, Riverside, CA 92521-0124, USA }%
\author{Zhen Jin}%
 \altaffiliation[]{Corresponding author}\email{jinzhn@263.net}
\affiliation {%
Department of Mathematics, North University of China, Taiyuan,
Shan'xi 030051, People's Republic of China
}%
\date{\today}

\begin{abstract}
Alexander B. Medvinsky \emph{et al} [A. B. Medvinsky, I. A.
Tikhonova, R. R. Aliev, B.-L. Li, Z.-S. Lin, and H. Malchow, Phys.
Rev. E \textbf{64}, 021915 (2001)] and Marcus R. Garvie \emph{et al} [M. R. Garvie and C. 
Trenchea, SIAM J. Control. Optim. \textbf{46}, 775-791 (2007)] shown that the minimal
spatially extended reaction-diffusion model of phytoplankton-zooplankton can exhibit
both regular, chaotic behavior, and spatiotemporal patterns in a
patchy environment. Based on that, the spatial plankton model is
furtherly investigated by means of computer simulations and theoretical analysis in the present paper when its parameters would be expected in the case of mixed Turing-Hopf bifurcation region. Our results show that the spiral
waves exist in that region and the spatiotemporal chaos emerge, which 
arise from the far-field breakup of the spiral
waves over large ranges of diffusion coefficients of phytoplankton
and zooplankton.  Moreover, the 
spatiotemporal chaos arising from the far-field breakup of spiral waves does not gradually involve the whole space within
that region. Our results are confirmed by means of computation spectra and nonlinear bifurcation of wave trains.
Finally, we give some explanations about the spatially 
structured patterns from the community level.

\end{abstract}


\maketitle
\section{Introduction}

There is a growing interest in the spatial pattern dynamics of
ecological
systems~\cite{amritkar:258102,pekalski:021913,Sayama2003,gilad:098105,mobilia2006,mobilia2005,
Blasius1999,PhysRevLett.87.198101,PhysRevE.67.056602,PhysRevE.64.021915,medvinsky:311,Gurney1998,Murray2002}.
However, many mechanisms of the spatio-temporal variability of
natural plankton populations are not known yet. Pronounced physical
patterns like thermoclines, upwelling, fronts and eddies often set
the frame for the biological process. Measurements of the underwater
light field are made with state-of-the-art instruments and used to
calculate concentrations of phytoplankton biomass (as chlorophyll)
as well as other forms of organic matter. Very high diffusion of the
marine environment would prevent the formation of any stable patch
spatial distribution with much longer life-time than the typical
time of biodynamics. Meanwhile, in addition to very changeable
transient spatial patterns,  there also exist other spatial patterns
in marine environment, much more stable spatial structure associated
with ocean fronts, spatiotemporal
chaos~\cite{medvinsky:311,PhysRevE.64.021915,Petrovskii}, cyclonic rings, and
so called meddies~\cite{Laurence}. In fact, it is significant to
create the biological basis for understanding spatial patterns of
plankton~\cite{EsaRanta11281997}. For instance, the impact of space on the persistence
of enriched ecological systems was proved in laboratory experiments~\cite{Luckinbill}. Recently, it has been shown both in laboratory experiments~\cite{Holyoak} and theoretically~\cite{Jansen,Jansen2000,Jansen2001,Petrovskii} that the existence of a spatial structure makes a predator-prey system less prone to extinction. This is due to the temporal variations of the density of different sub-populations can become asynchronous and the events of local extinction can be compensated due to re-colonization from other sites in the space~\cite{Allen2001}.
 During a long period of time, all
the spiral waves have been widely observed in diverse physical,
chemical, and biological
systems~\cite{RevModPhys.65.851,PhysRevLett.87.068101,Sawai2005,winfree:303}.
However, a quite limited number of
documents~\cite{Gurney1998,medvinsky:311,Biktashev,Garvie2007,garvie:775}
concern the spiral wave pattern and its breakup in the ecological systems.

The investigation of transition from regular patterns to
spatiotemporally chaotic dynamics in spatially extended systems remains a
challenge in nonlinear
science~\cite{Markus2004,PhysRevLett.82.1160,RevModPhys.65.851,Petrovskii}. In
a nonlinear ecology system, the two most commonly seen patterns are
spiral waves and turbulence (spatio-temporal chaos) for the level of the community~\cite{Craigjohnson2002}. It has been recently shown that spontaneous spatiatemoporal
pattern formation is an instrinsic property of a predator-prey system~\cite{Pascual1993,Sherratt1997,Petrovskii2001,medvinsky:311,Petrovskii,Sherratt2001} and  spatiotemporal structures play an important role in
ecological systems. For example, spatially induced speciation
prevents the extinction of the predator-prey
models~\cite{Savill1998,Gurney1998,medvinsky:311}. So far, plankton patchiness has been observed on a wide range of spatial temporal scales~\cite{Abraham1998,Folt1999}. There
exist various, often heuristic explanations of the spatial patterns
phenomenon for these systems. It should be noted that, although conclusive
evidence of ecological chaos is still to be found, there is a growing number of indications of chaos in real ecosystems~\cite{MartenScheffer01011991,Hanski1993,Ellner1995,Dennis2001}.

Recently developed models show that spatial self-structuring in multispecies systems can meet both criteria and provide a rich substrate for community-level section and a major transition in evolution. In present paper, the scenario in the spatially extended 
plankton ecological system is observed by means of the numerical simulation. The system has been demonstrated to exhibit regular or chaostic, depending on the initial conditions and the parameter values~\cite{PhysRevE.64.021915,garvie:775}.
We find that the far-field breakup of the spiral wave leads to complex
spatiotemporal chaos (or a turbulentlike state) in the
spatially extended plankton model~\eqref{eq:II2}. Our results show that regular spiral wave pattern shifts into spatiotemporal chaos pattern by modulating the diffusion coefficients of the species.

\section{model}

In this paper we study the spatially extended nutrient-phytoplankton-zooplankton-fish reaction-diffusion system.
Following Scheffer's minimal approach~\cite{Scheffer1991a}, which was originally formulated as a system of ordinary diffential equation (ODEs) and later developed models~\cite{PhysRevE.64.021915,medvinsky:311,Malchow1993,MercedesPascual,garvie:775},
as a further investigation, we study a two-variable
phytoplankton and zooplankton model on the level of the community to describe pattern formation with the diffusion. The dimensionless model is written as
\begin{subequations}
\label{eq:II2}
\begin{equation}
\frac{\partial p}{\partial t}=r
   p(1-p)-\frac{ap}{1+bp}h+d_{p}\nabla^{2} p
 ,\label{subeq:II2a}
\end{equation}
\begin{eqnarray}
\frac{\partial h}{\partial
t}=\frac{ap}{1+bp}h-mh-f\frac{nh^2}{n^2+h^2}+d_{h}\nabla^{2}
   h,\label{subeq:II2b}
\end{eqnarray}
\end{subequations}
where the parameters are $r$, $a$, $b$, $m$, $n$, $d_{p}$, $d_{h}$, and
$f$ which refer to work in Refs.~\cite{PhysRevE.64.021915,medvinsky:311}. The explanation of model~\eqref{eq:II2} relates to the nutrient-phytoplankton-zooplankton-fish ecological system [see Refs.~\cite{garvie:775,Scheffer1991a,PhysRevE.64.021915} for details]. The local dynamics are given by
  \begin{subequations}
\label{eq:II3}
\begin{equation}
g_{1}(p,h)=r
   p(1-p)-\frac{ap}{1+bp}h,\label{subeq:II3a}
\end{equation}
\begin{eqnarray}
g_{2}(p,h)=\frac{ap}{1+bp}h-mh-f\frac{nh^2}{n^2+h^2}.\label{subeq:II3b}
\end{eqnarray}
\end{subequations}

\begin{figure}[htp]
\includegraphics[width=7cm]{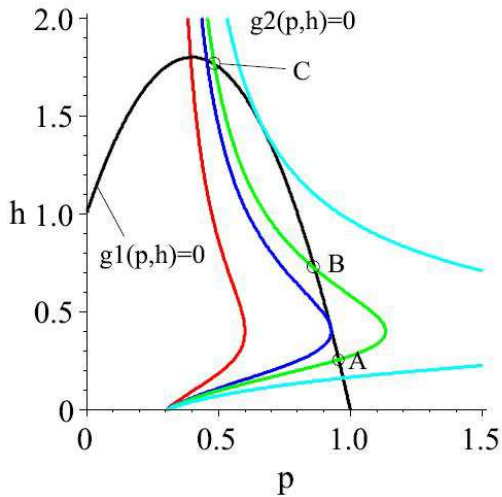}
\caption{\label{fig1}The sketch map for the bistability and the Hopf
bifurcation in the system~\eqref{eq:II3} with $r=5.0$, $a=5.0$,
$b=5.0$, $m=0.6$, and $n=0.4$.  The black curve is the $g_{1}(p,h)$.
The colored curves are $g_{2}(p,h)$ with different values of $f$.
The red curve: $f=0.3$; the blue: $f=0.445$; the green: $f=0.5$; and
the cyan: $f=0.658$.}
\end{figure}

\begin{figure}[htp]
\includegraphics[width=8cm]{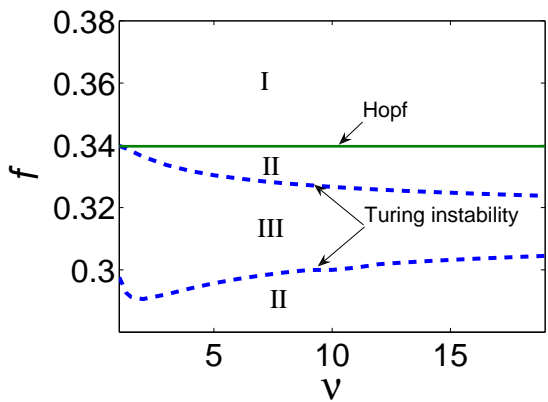}
\caption{\label{Bifurcationdiagram} The sketch map of parameter
space $(f,\nu)$ bifurcation diagrams for the spatially extended system~\eqref{eq:II2}
with $r=5.0$, $a=5.0$, $b=5.0$, $m=0.6$, $d_{p}=0.05$, and $n=0.4$.}
\end{figure}

From the earlier
results~\cite{Malchow1993} about non-spatial system of model~\eqref{eq:II2} by means of numerical bifurcation
analysis show that the bifurcation and bistability can be found
in the system~\eqref{eq:II2} when the parameters are varied within a
realistic range. For the fixed parameters (see the caption of
Fig.~\ref{fig1} and \ref{Bifurcationdiagram}), we can see that the $f$ controls the distance from
Hopf bifurcation.  For larger $f$, there exists only one stable
steady state. As $f$ is decreased further, the homogeneous steady state undergoes a saddle
node bifurcation (SN), that is $f_{SN}=0.658$. In this case, a stable and an
unstable steady state become existence. Moreover, the bistability will emerge
when the parameter $f$ lies the interval $f_{SN}>f>f_{c}=0.445$
(this value is more than the Hopf onset, $f_{H}=0.3397$). There are
three steady states: with these kinetics A and C are linearly stable
while B is unstable. Outside this interval, the system~\eqref{eq:II2}
has unique nontrivial equilibrium. Recent studies~\cite{medvinsky:311,garvie:775} shown that the systems~\eqref{eq:II2} can
well-develop the spiral waves in the oscillation regime, but where the
authors only consider the special case, i.e., $d_p=d_h$. A few important issue have not yet
been properly addressed such as the spatial pattern if $d_p\neq d_h$. 

Here we report the result that emergence of spatiotemporal chaos due to
breakup in the system under the $d_h\neq d_p$ case. We may now use
the $f$ and diffusion ratio, $\nu=d_{h}/d_{p}$, as control parameters to evaluate the
region for the spiral wave. Turing instability in reaction-diffusion can be recast in
 terms of matrix stability~\cite{Scheel2003,Satnoianu}. Such with the help of Maple software
assistance algebra computing, we obtain the  parameters space
$(f,\nu)$ bifurcation diagrams of the spiral waves as showing
Fig.~\ref{Bifurcationdiagram}, in which two lines are plotted, Hopf
line (solid) and Turing lines (dotted) respectively. In
domain I, located above all three bifurcation lines, the homogeneous steady states is the only 
stable solution of the system. Domain II are 
regions of homogeneous oscillation in two
dimensional spaces~\cite{liuqx2007}. In domain III, both Hopf and Turing instabilities occur, (i.e., mixed Turing-Hopf modes arise), in which the system generally
produces the phase waves. Our results show that the
system has spiral wave in this regions. One can see that a Hopf bifurcation can occur at the steady when the parameter $f$ passes through a critical values $f_H$ while the diffusion coefficients $d_p=d_h=0$ and the bifurcation periodic solutions are stable. From our analysis (see Fig.~\ref{Bifurcationdiagram}), one 
could also see that the diffusion can induce Turing type instability for the spatial homogeneous stable periodic solutions and the spatially extended model~\eqref{eq:II2} exhibit spatio-temporal chaos patterns. These spatial pattern formation arise from interaction between Hopf and Turing modes, and their subharmonics near hte codimension-two Hopf-Turing bifucation point. Special, it is interesting that spiral wave and travelling wave will appear when the parameters correspond to the Turing-Hopf bifurcation region III in the spatially extended model~\eqref{eq:II2}, i.e., the Turing instability and Hopf bifurcation occur simultaneously. 

\section{Numerical results}

 The simulation is done in a two-dimensional (2D) Cartesian
coordinate system with a grid size of $600\times 600$. The fourth
order Runger-Kutta integrating method is applied with a time step
$\Delta t=0.005$ time unit and a space step $\Delta x=\Delta y=0.20$
length unit. The results remain the same when the reaction-diffusion equations
were solved numerically in one and two spatial dimensions using a finite-difference approximation for the spatial derivatives and an explicit Euler method for the time integration. Neumann (zero-flux) boundary conditions were emmployed in our simulation. The diffusion terms in Eqs.~\eqref{subeq:II2a} and
\eqref{subeq:II2b} often describe the spatial mixing of species due
to self-motion of the organism. The typical diffusion coefficient of plankton patterns $d_{p}$ is about $0.05$, based on the parameters estimatie of Refs~\cite{Jorgensen,Okubo1980} using the relationship between
turbulent diffusion and the scale of the space in the
sea. In the previous studies~\cite{PhysRevE.64.021915,medvinsky:311,Malchow1993,MercedesPascual,garvie:775}, the authors provided a valueable insight into the role of spatial pattern for the system~\eqref{eq:II2} if $d_p=d_h$. From
the biological meaning, the diffusion coefficients should satisfy
$d_{h}\geq d_{p}$. However, in nature waters it is turbulent
diffusion that is supposed to dominate plankton
mixing~\cite{Sugihara}, when $d_{h}<d_{p}$ is allowed.  The other
reason for choosing such parameter is that it is well-known new
patterns, such as Turing patterns, can emerge in reaction-diffusion
systems in which there is an imbalance between the diffusion
coefficients $d_p$ and $d_h$~\cite{Turing1952,RevModPhys.65.851}.
Therefore, we set $\nu=d_{h}/d_{p}$, and investigated whether a
spiral wave would break up into complex spatiotemporal chaos when
the diffusion ratio was varied. Throughout this paper, we fix
$d_{p}=0.05$ and $d_{h}$ is a control parameter.

\begin{figure}[htp]
\hspace{-0.7cm}(A)
\includegraphics[width=2.5cm]{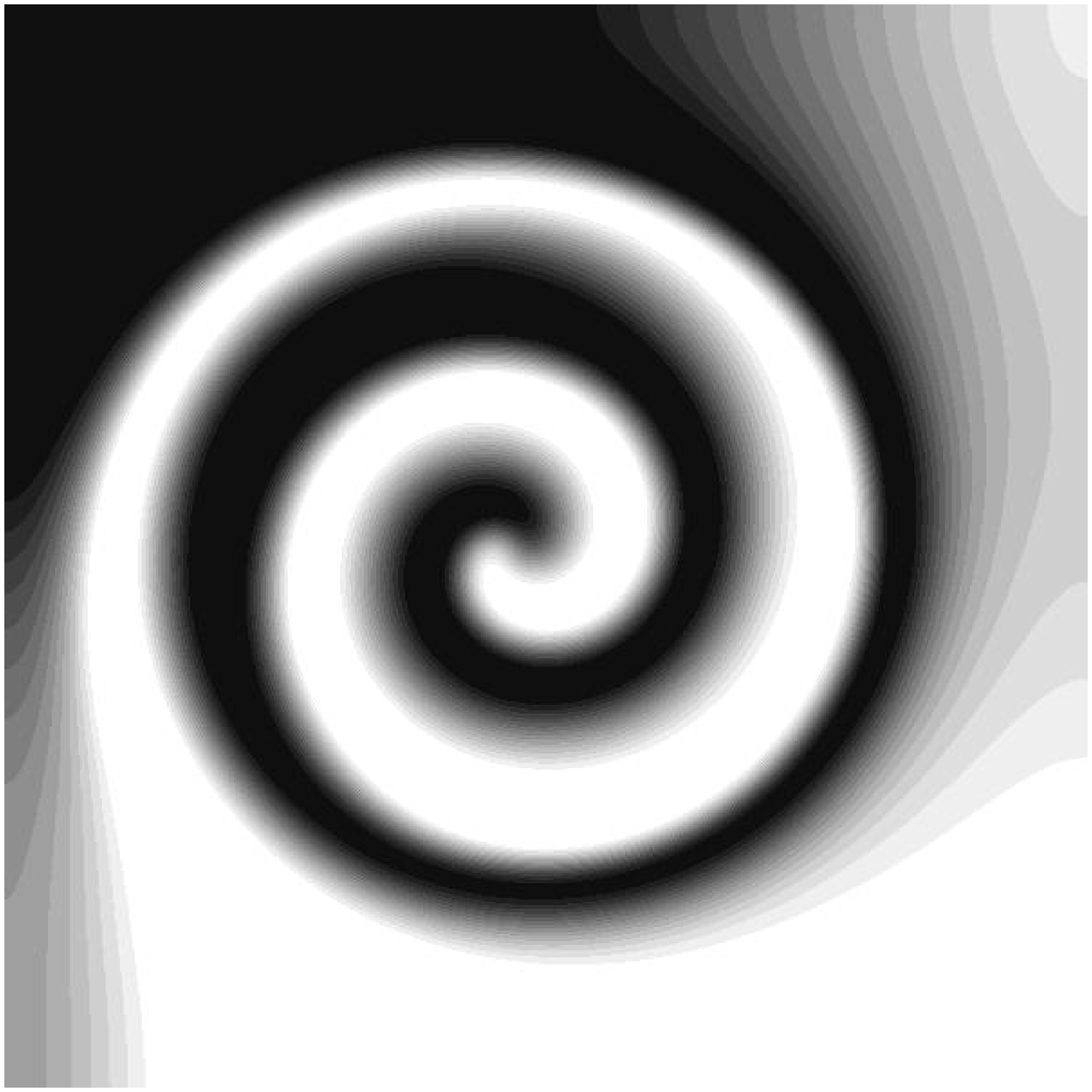}
\includegraphics[width=2.5cm]{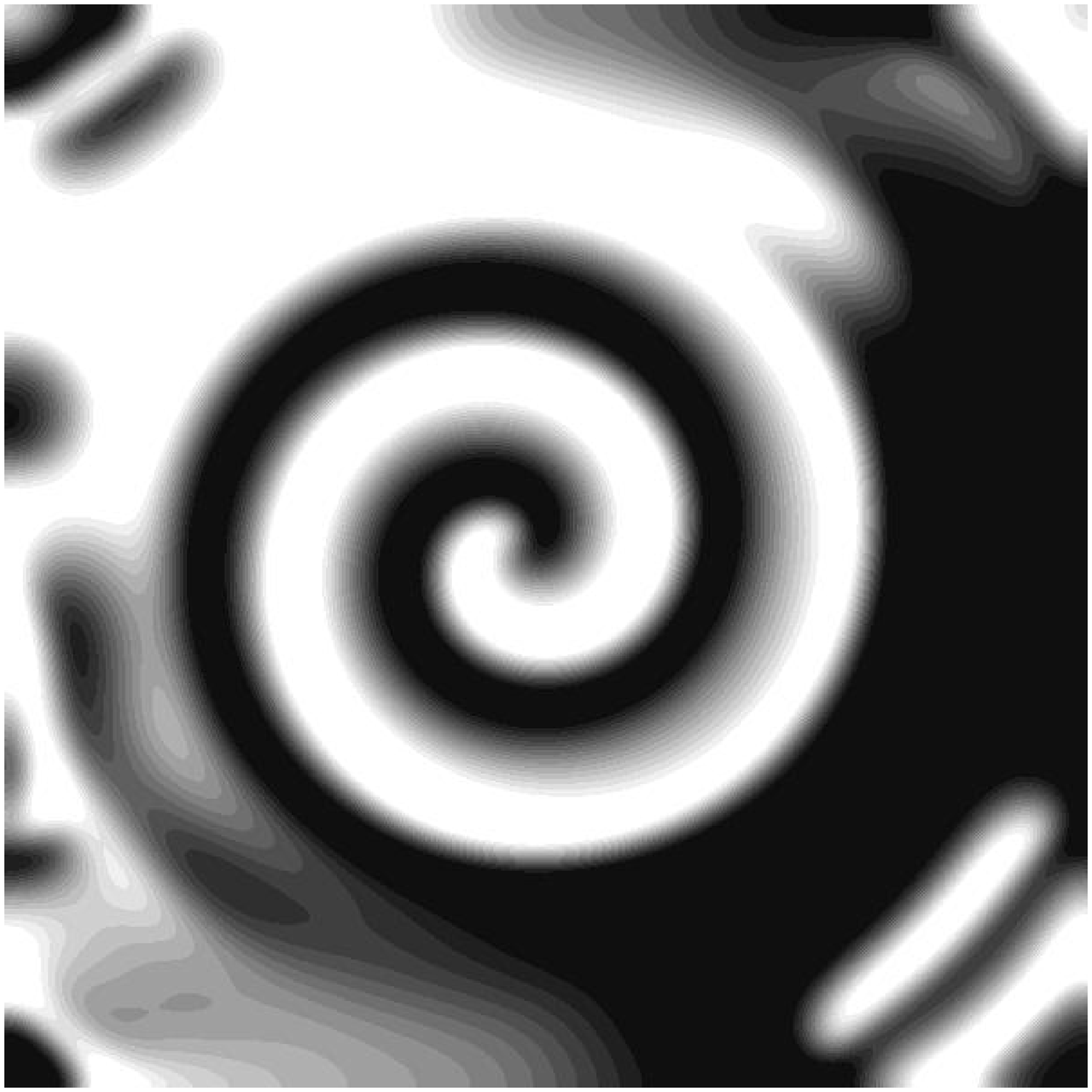}
\includegraphics[width=2.5cm]{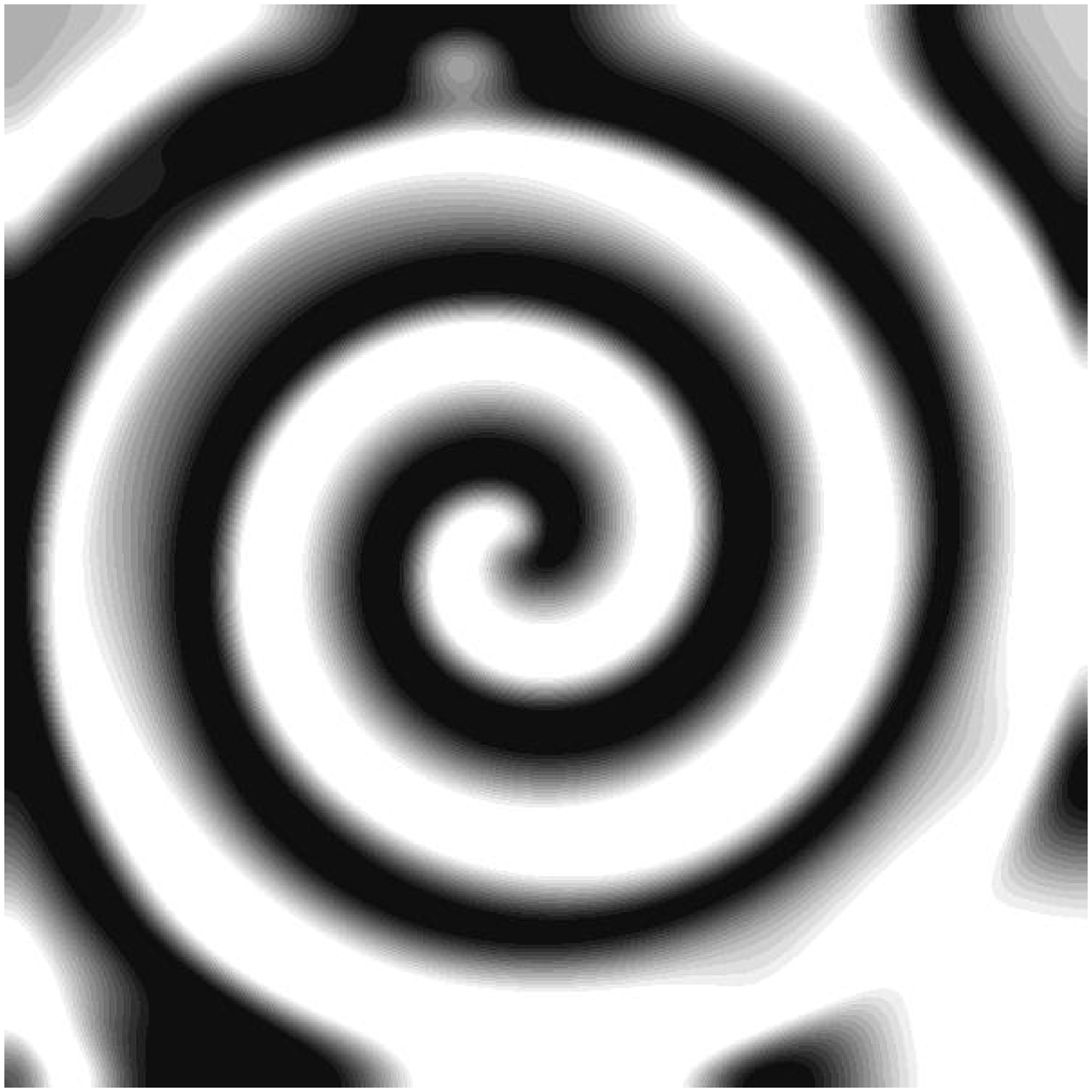}\\
\includegraphics[width=2.5cm]{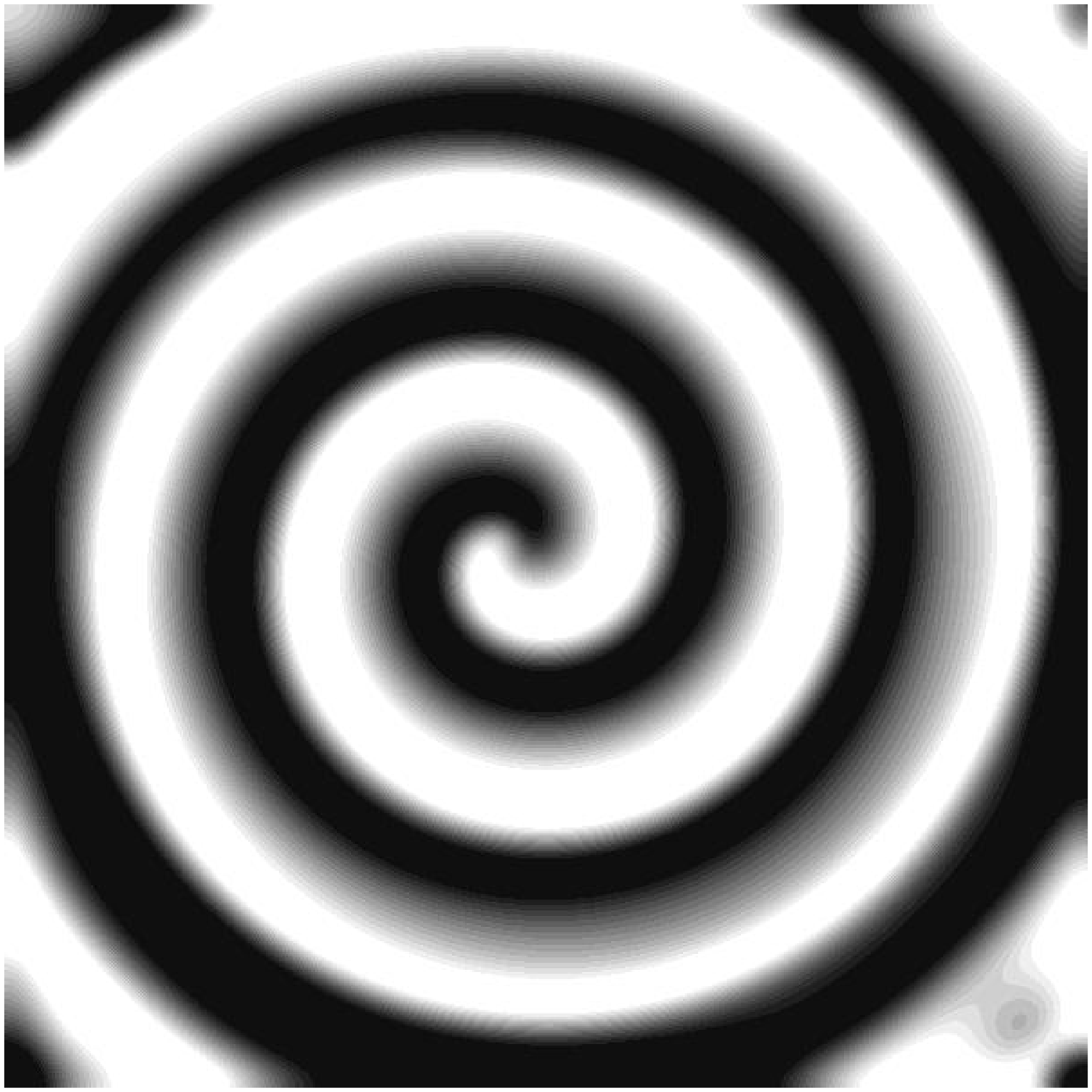}
\includegraphics[width=2.5cm]{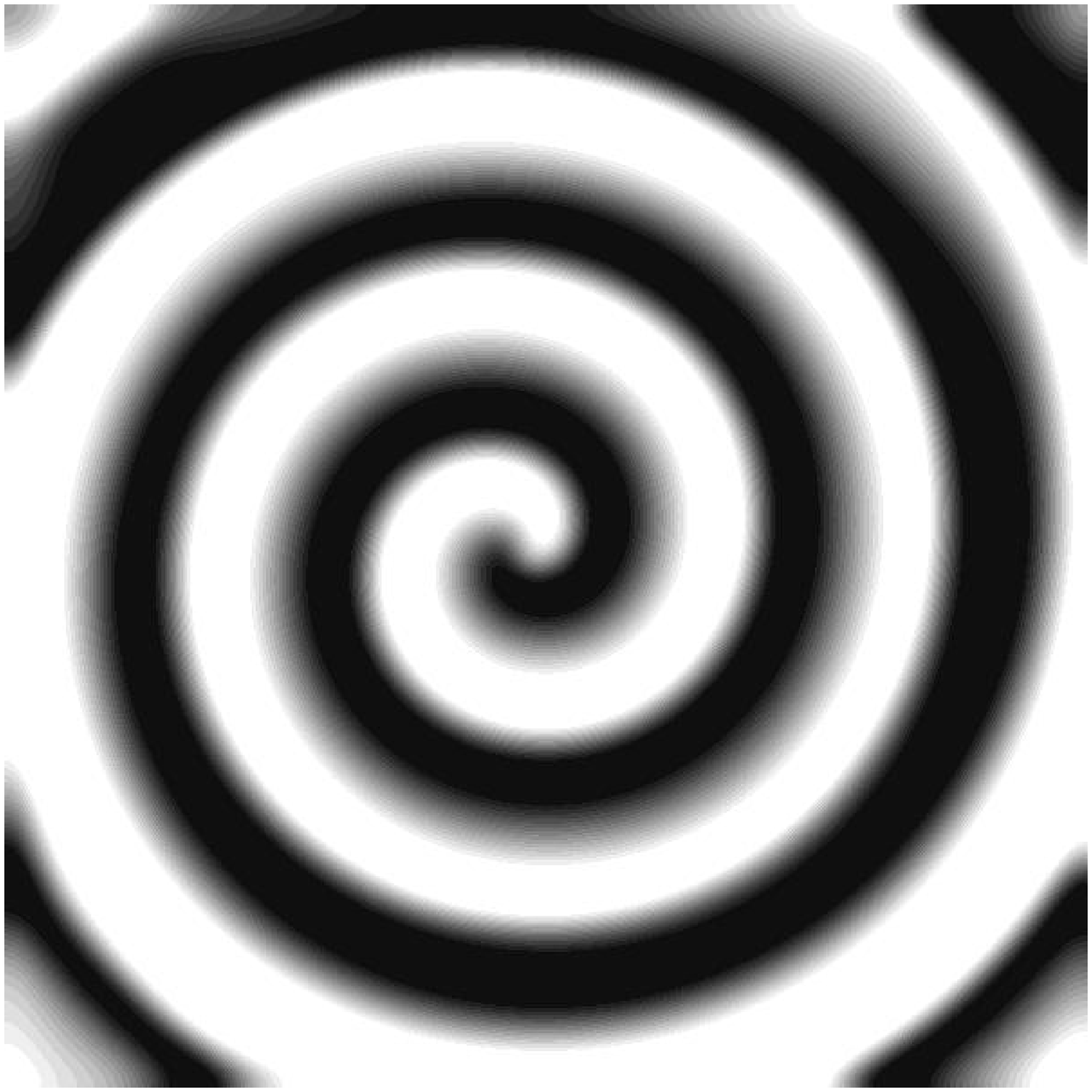}
\includegraphics[width=2.5cm]{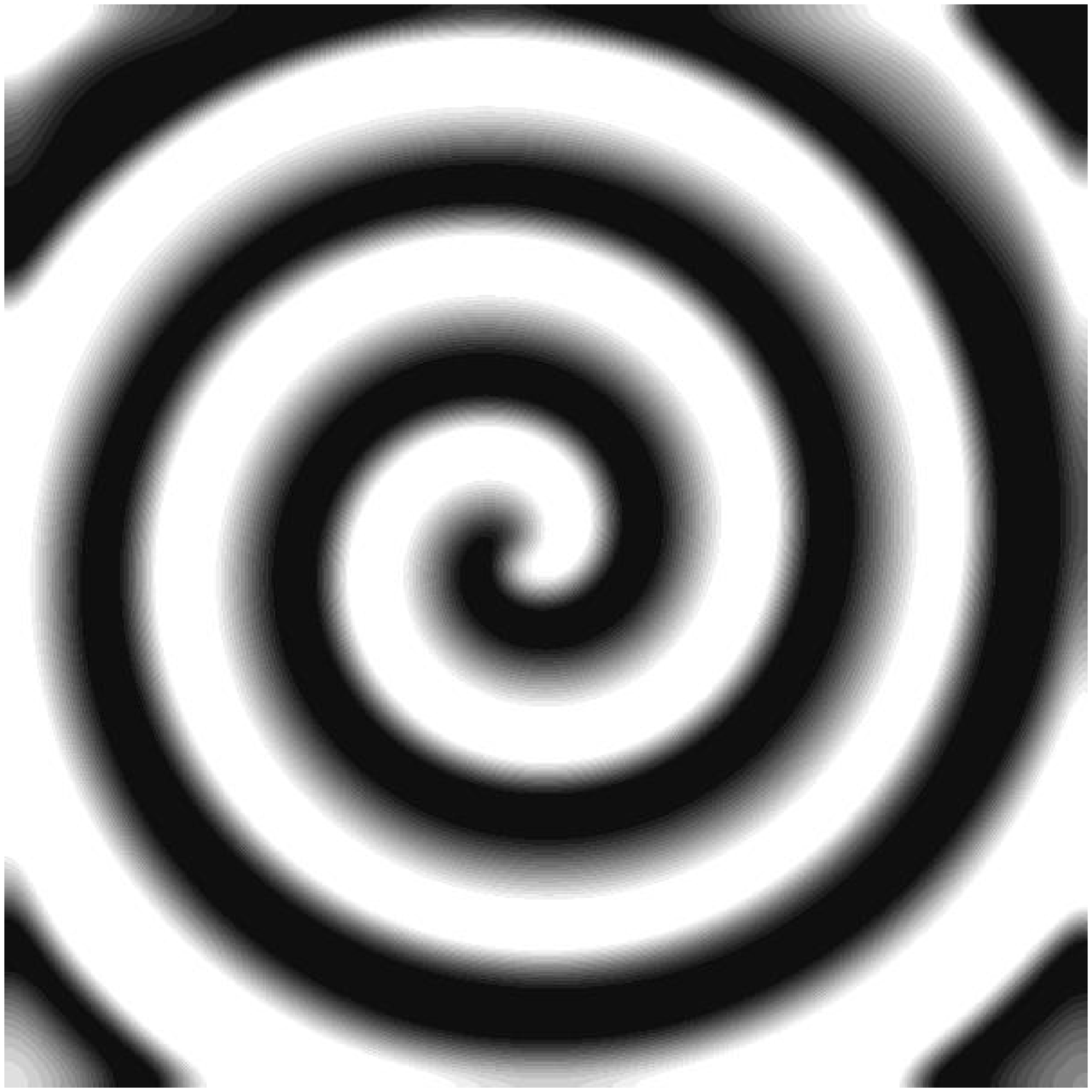}\\
\hspace{-0.7cm}(B)
\includegraphics[width=2.5cm]{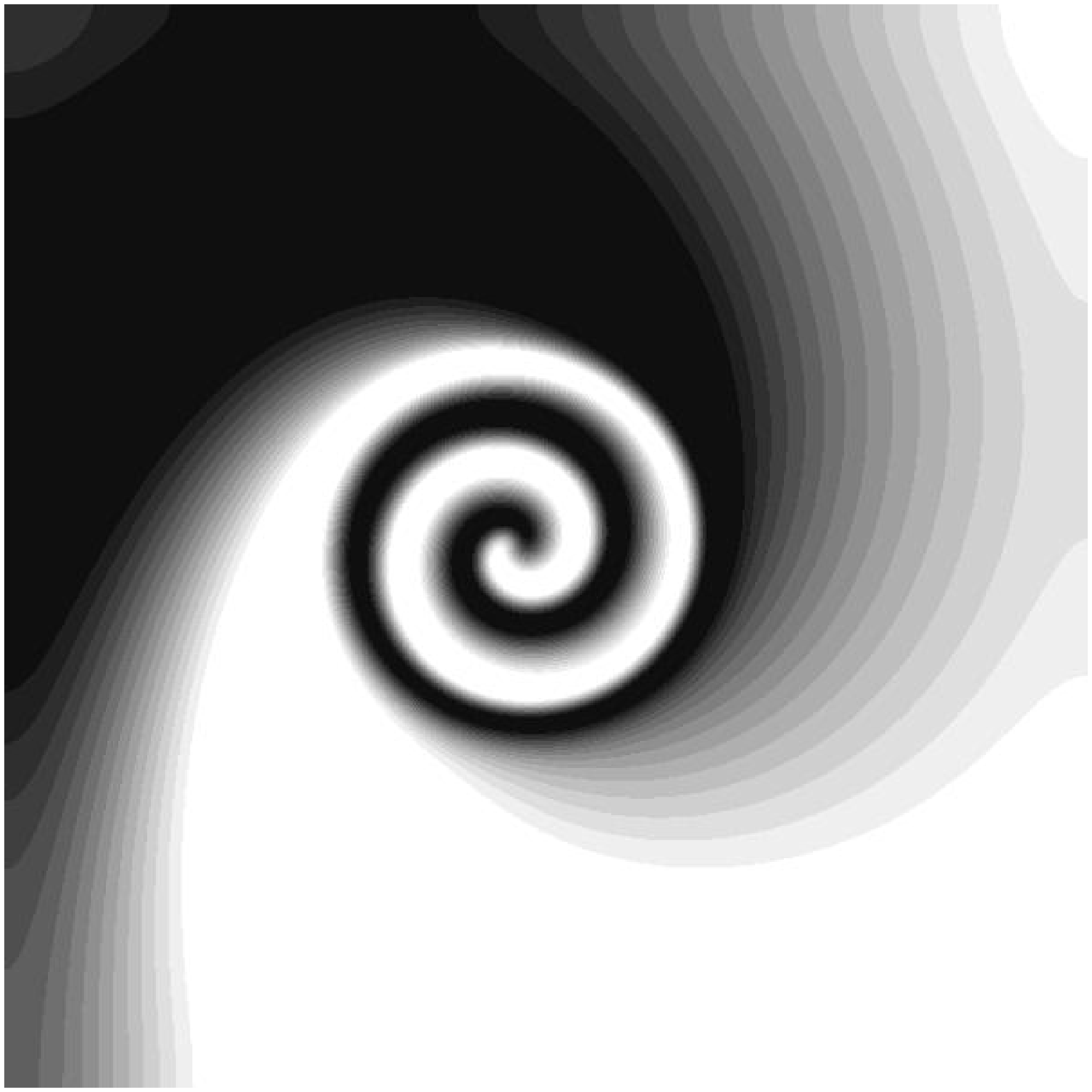}
\includegraphics[width=2.5cm]{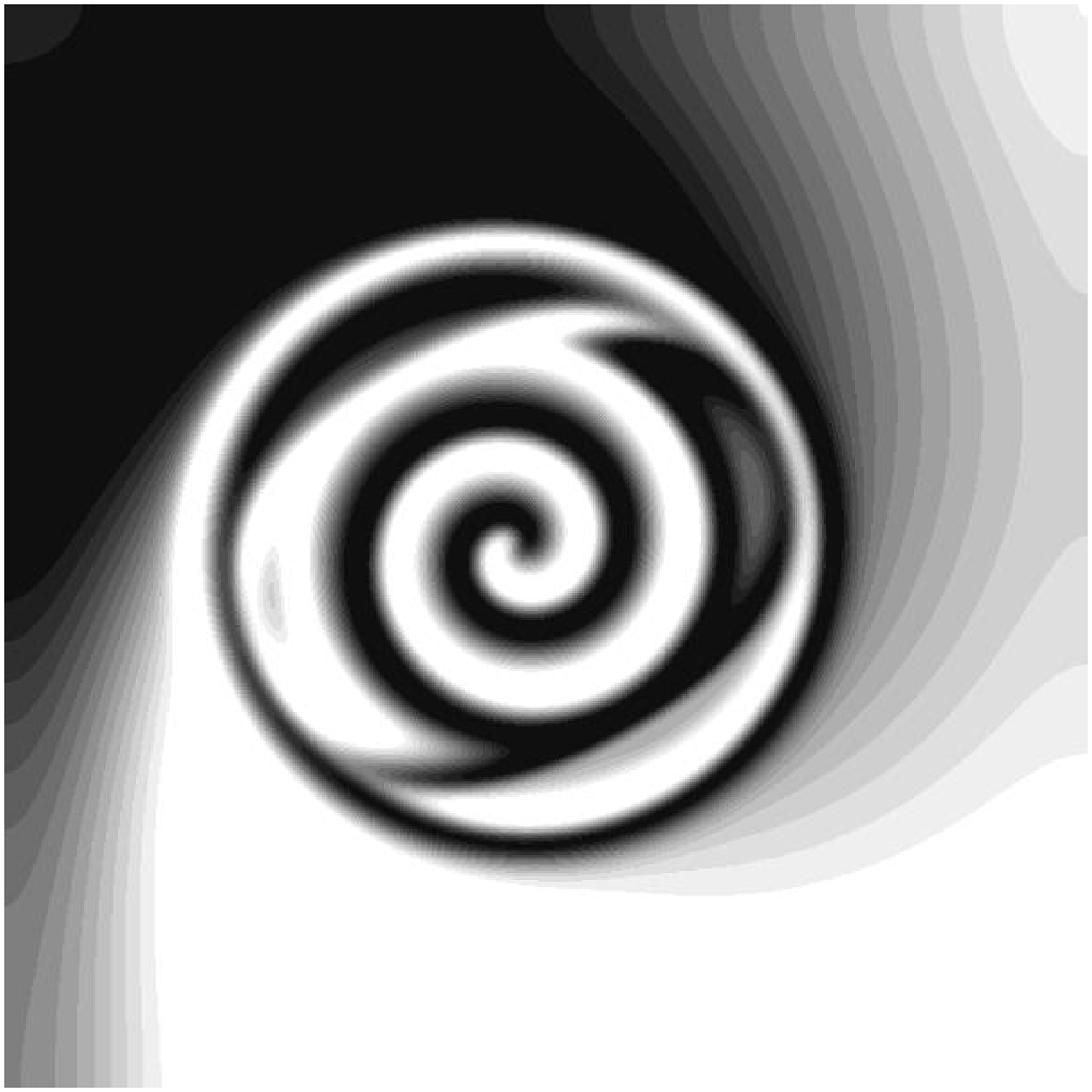}
\includegraphics[width=2.5cm]{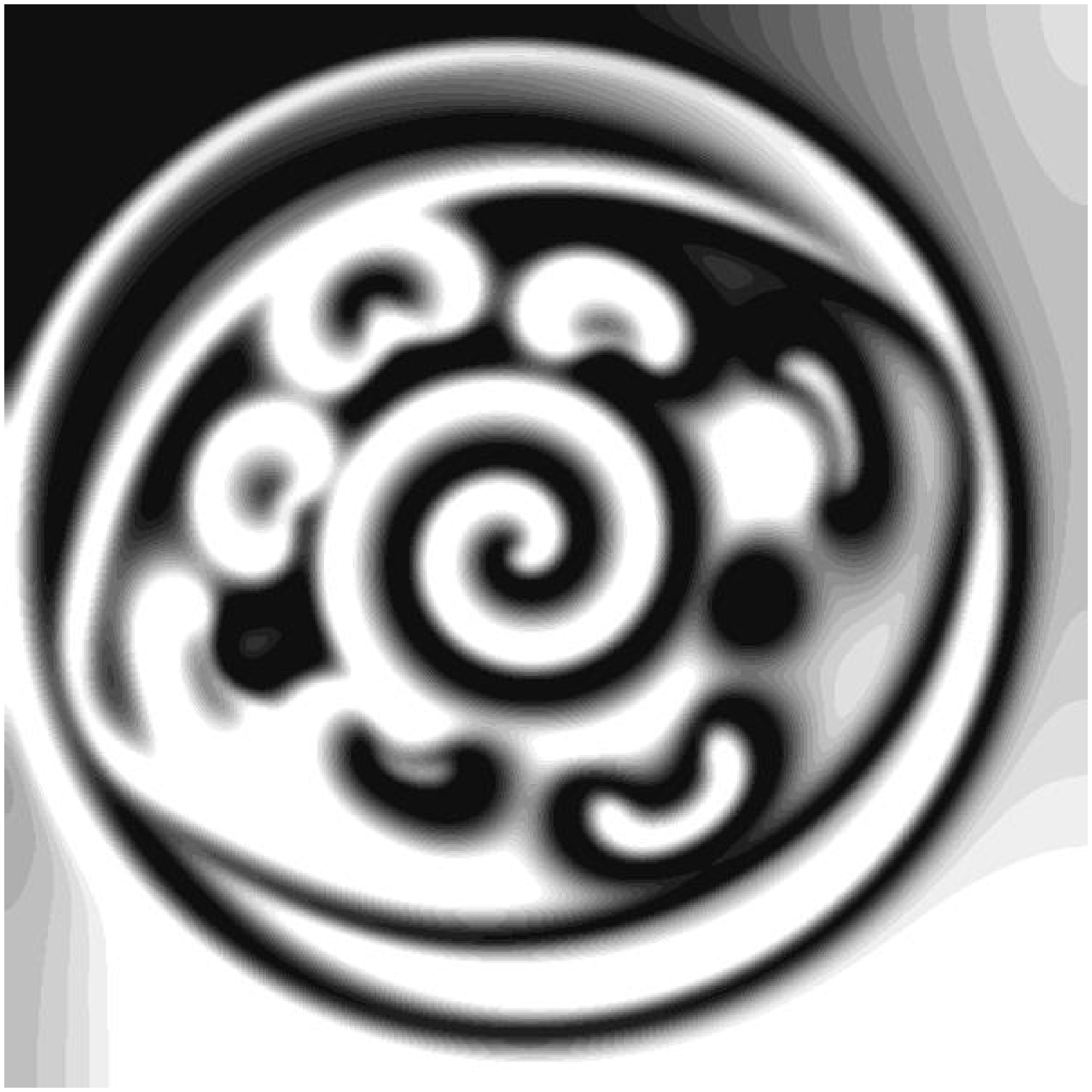}\\
\includegraphics[width=2.5cm]{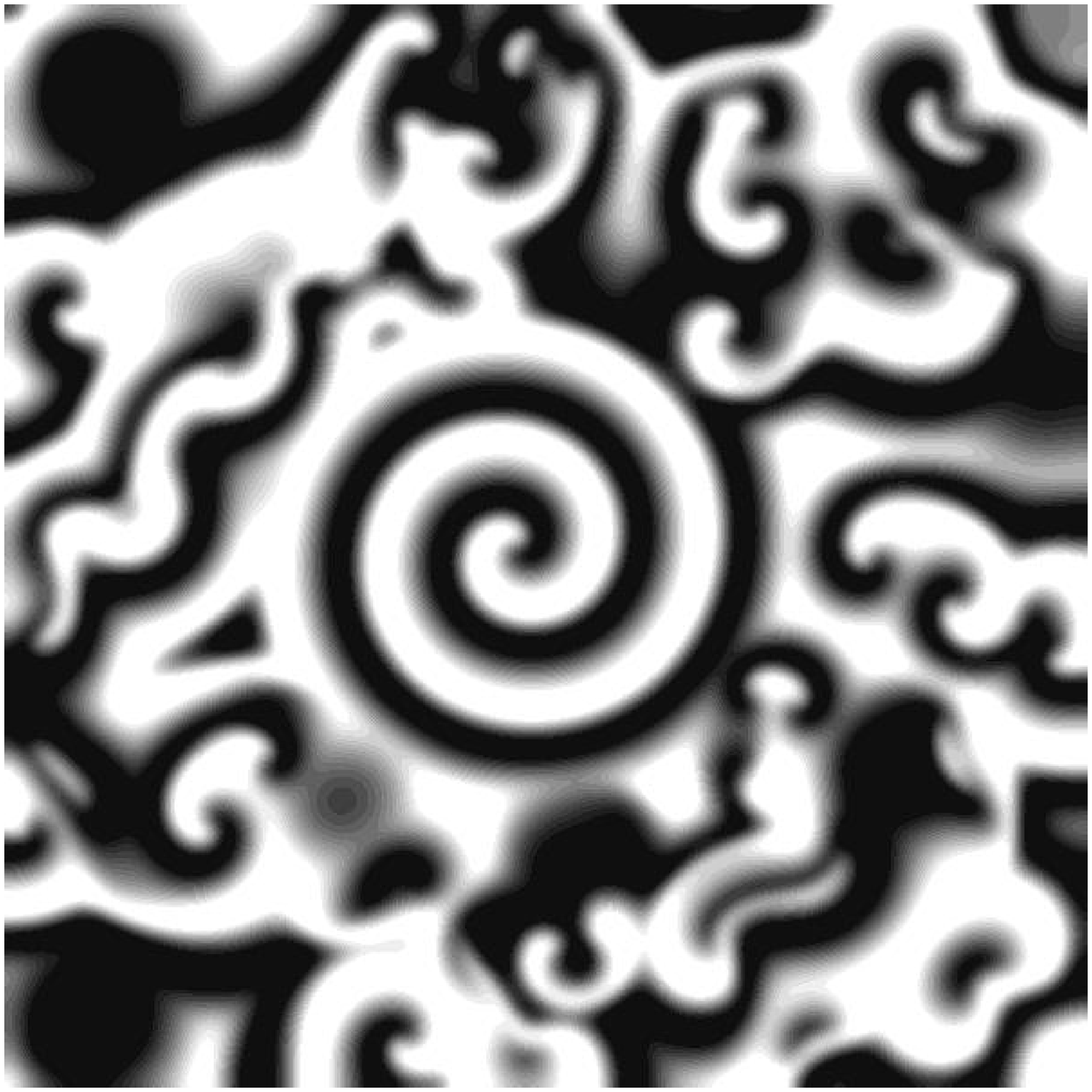}
\includegraphics[width=2.5cm]{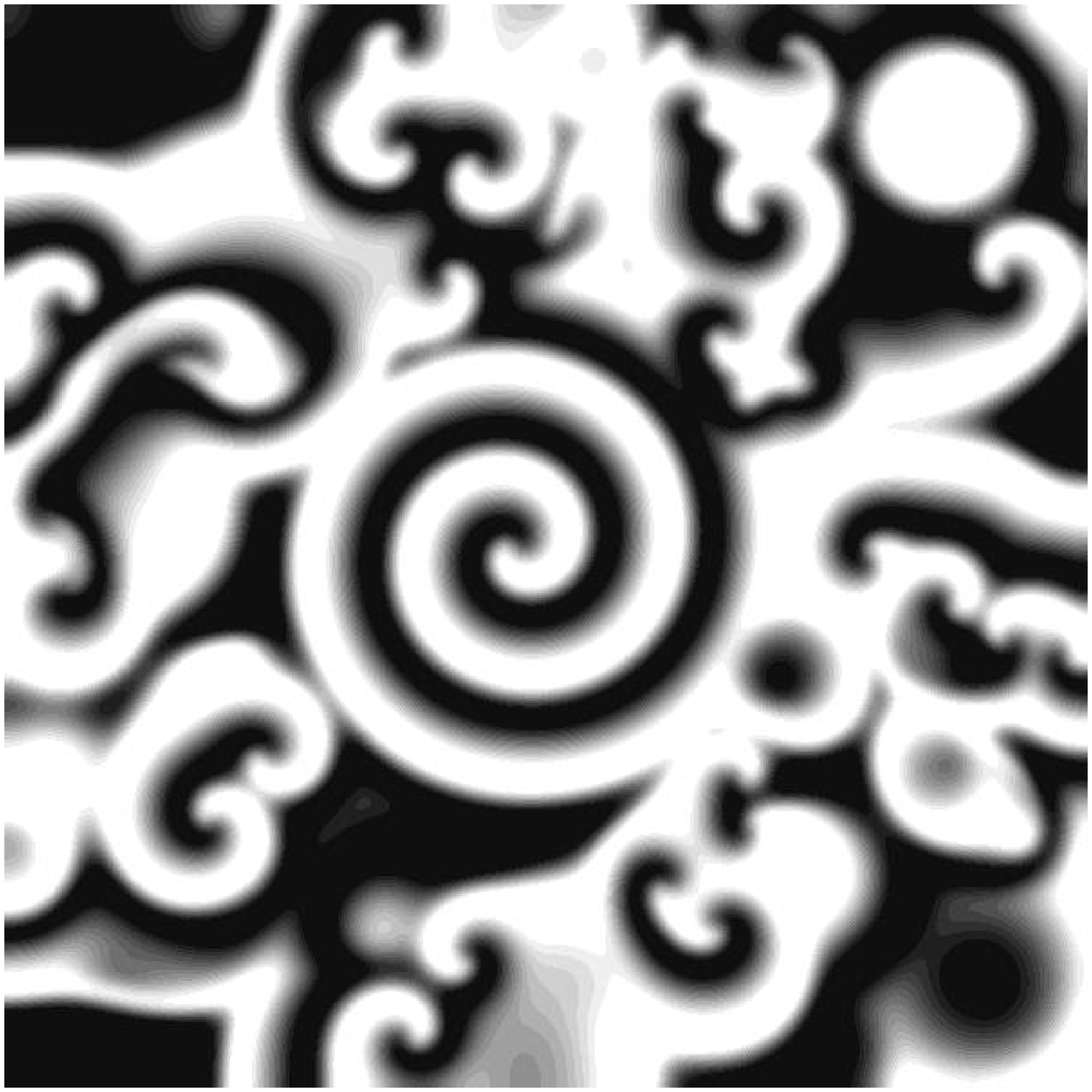}
\includegraphics[width=2.5cm]{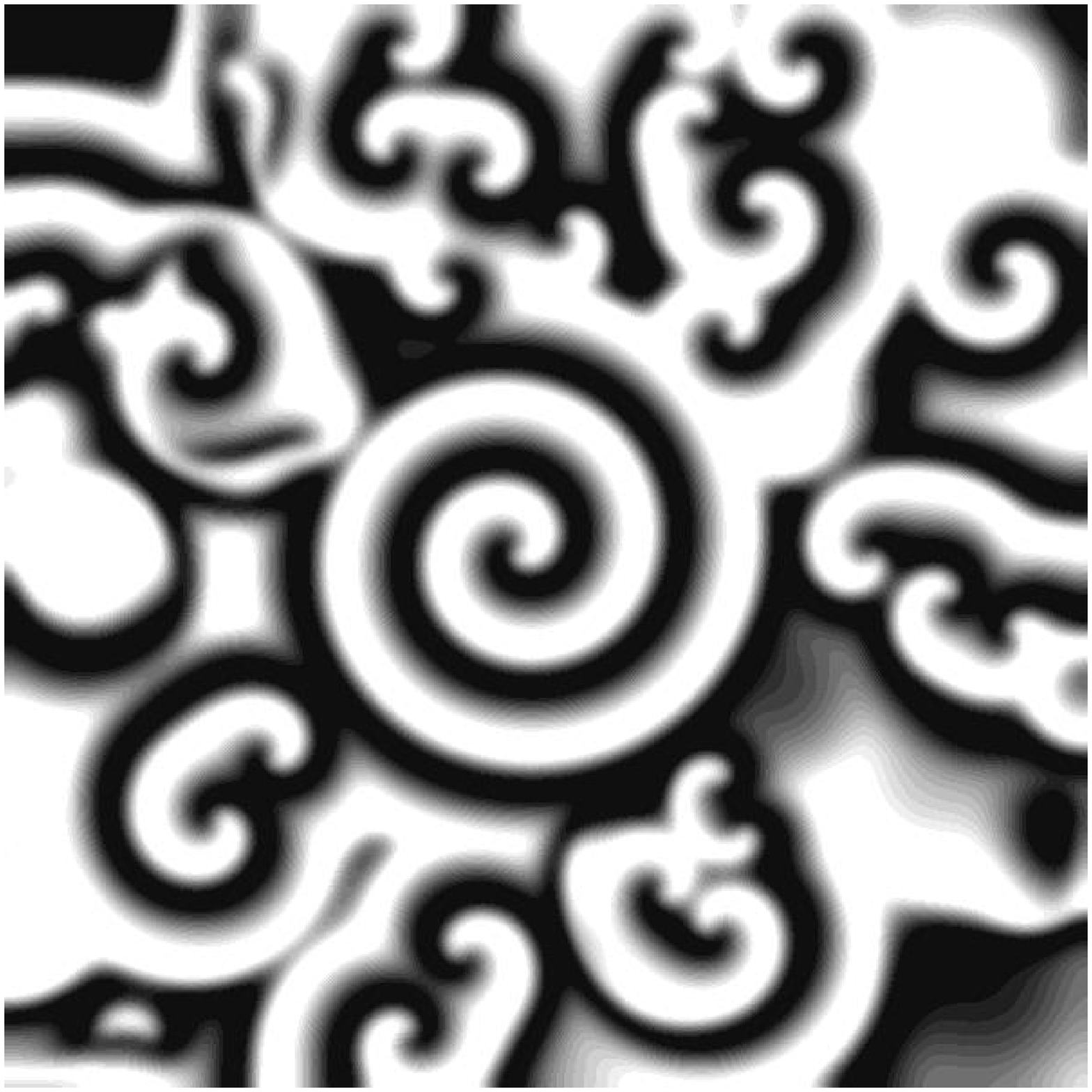}\\
\includegraphics[width=7cm]{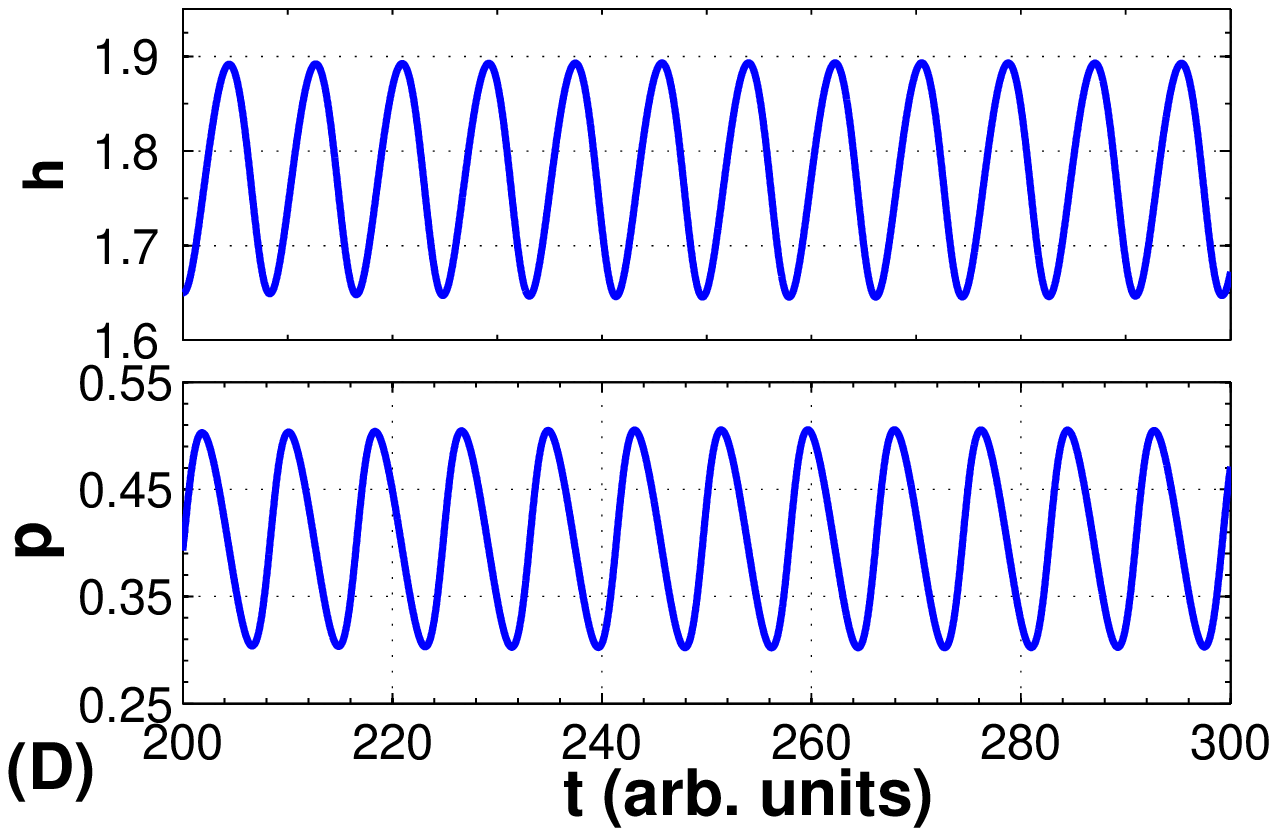}
\caption{\label{spiral} Well developed spiral waves and some
properties of them. The figures show simulations of the
system~\eqref{eq:II2} with $r=5$, $a=5$, $b=5$, $m=0.6$, $n=0.4$,
$d_{p}=0.05$, and $f=0.3$. (A) Well developed spiral waves shown at
subsequent snapshot in time, $d_{h}=0.2$. (B) Far-field breakup of
the spiral waves shown at subsequent snapshot in time, $d_{h}=0.002$.
The white (black) areas correspond to maximum (minimum) values of
$p$ [Additional movie format available from Ref.~\cite{movie1}]. (C)
Oscillations of the variable $p$ and $h$ at an arbitrary spatial
point within the regular spiral wave region for both scenarios. Each
figure is ran the long time until it spatial patterns are unchange.}
\end{figure}

In the following, we will show that the dynamic behavior of the
spiral wave qualitatively change as the control parameter $d_h$
increases from zero, i.e., the diffusion ratio $\nu$ increases from
zero, to more than one. For large $\nu$ ($\nu>1$), the outwardly
rotating spiral wave is completely stable everywhere, and fills in
the space when the proper parameters are chosen, as shown in
Fig.~\ref{spiral}(A). Figure~\ref{spiral}(A) shows a series of
snapshots of a well-developed single spiral wave formed
spontaneously for the variable $p$ in system~\eqref{eq:II2}. The
spiral is initiated on a $600\times 600$ grid by the cross-field
protocol (the initial distribution chosen in the form of allocated ``constant-gradient'' perturbation of the co-existence steady state) and zero boundary conditions are employed for simulations
in the two dimensions. From Fig.~\ref{spiral}(A) we can see that
the well-developed spiral waves are formed firstly by the evolution.
Inside the domain, new waves emerge, but are evolved by the spiral
wave growing from the center. The spiral wave can steadily grow
and finally prevail over the whole domain (a movie illustrating the
dynamical evolution for this case~\cite{movie1} [partly
$movie_{-}1$, $movie_{-}2$, and $movie_{-}3$ for $d_h=0.2$]).
Fig.~\ref{spiral}(B) shows that the spiral wave first break up far
away from the core center and eventually relatively large spiral
fragments are surrounded by a `turbulent' bath remain. The size of
the surviving part of the spiral does not shrink when $d_{h}$ is
further decreasing until finally $d_h$ equals to 0, which is
different from phenomenon that is observed  previous in the two-dimensional
space Belousov-Zhabotinsky and FitzHugn-Nagumo oscillatory
system~\cite{Markus2004,PhysRevLett.82.1160,xie:026107,
PhysRevE.62.7708,PhysRevLett.80.4811}, in which the breakup
gradually invaded the stable region near the core center, and
finally the spiral wave broke up in the whole medium.
Figure~\ref{spiral}(C) is the time sequences (arbitrary units) of
the variables $p$ and $h$ at an arbitrary spatial point within the
spiral wave region, from which we can see that the spiral waves are
caused by the accepted as ``phase waves" with substantially group
velocity, phase velocity and sinusoidal oscillation rather than the
relaxational oscillation with large amplitude. This breakup scenario
is similar to the breakup of rotating spiral waves observed in
numerical simulation in chemical
systems~\cite{Markus2004,PhysRevLett.82.1160,xie:026107,
PhysRevE.62.7708,PhysRevLett.80.4811}, and experiments in BZ systems~\cite{Ouyang1996,Ouyang2000}, which shows that spiral wave breakup
in these systems was related to the Eckhaus instability and more
important, the absolute instability.

\begin{figure}[h]
\includegraphics[width=7cm]{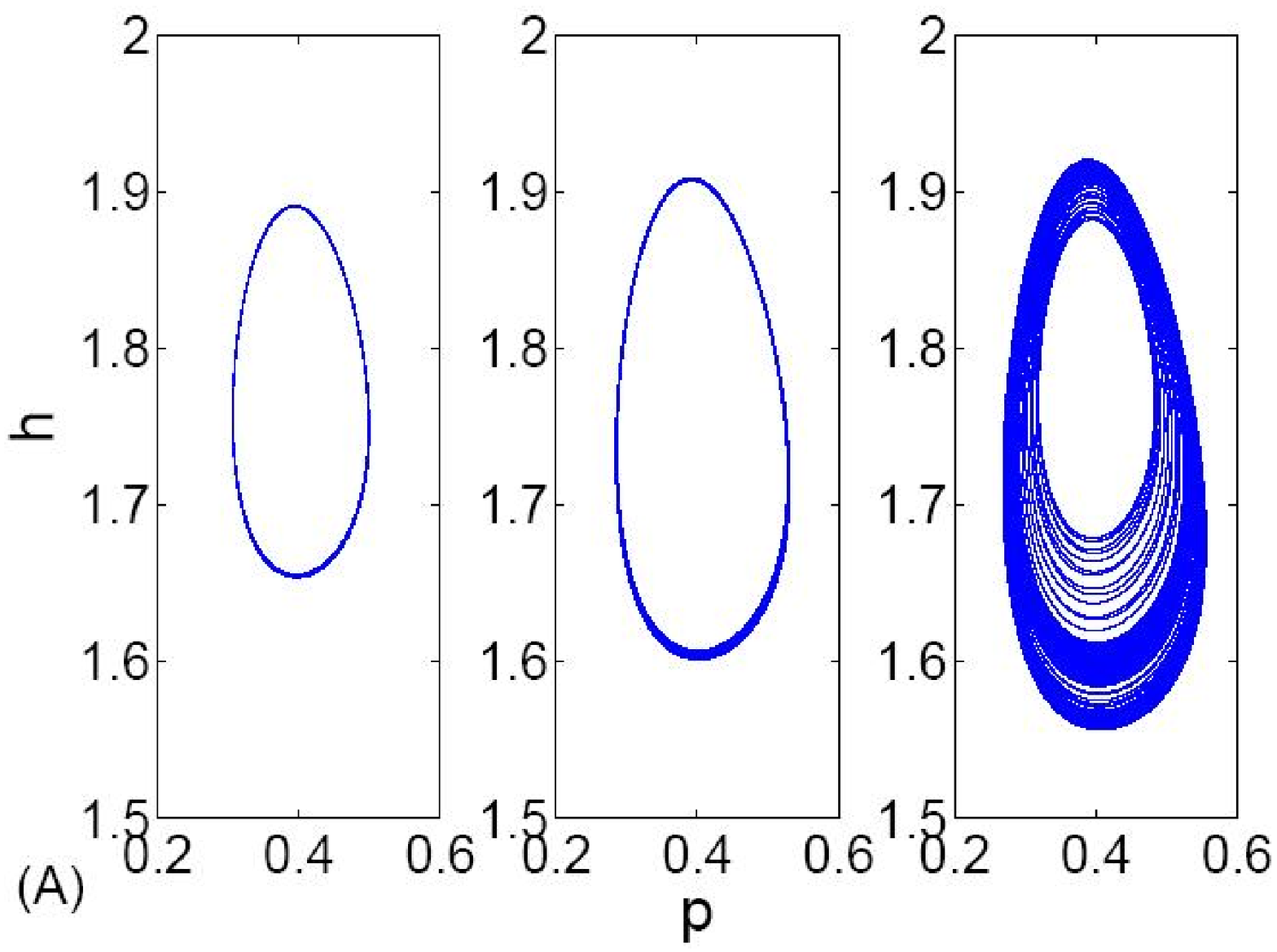}\\
\includegraphics[width=7cm]{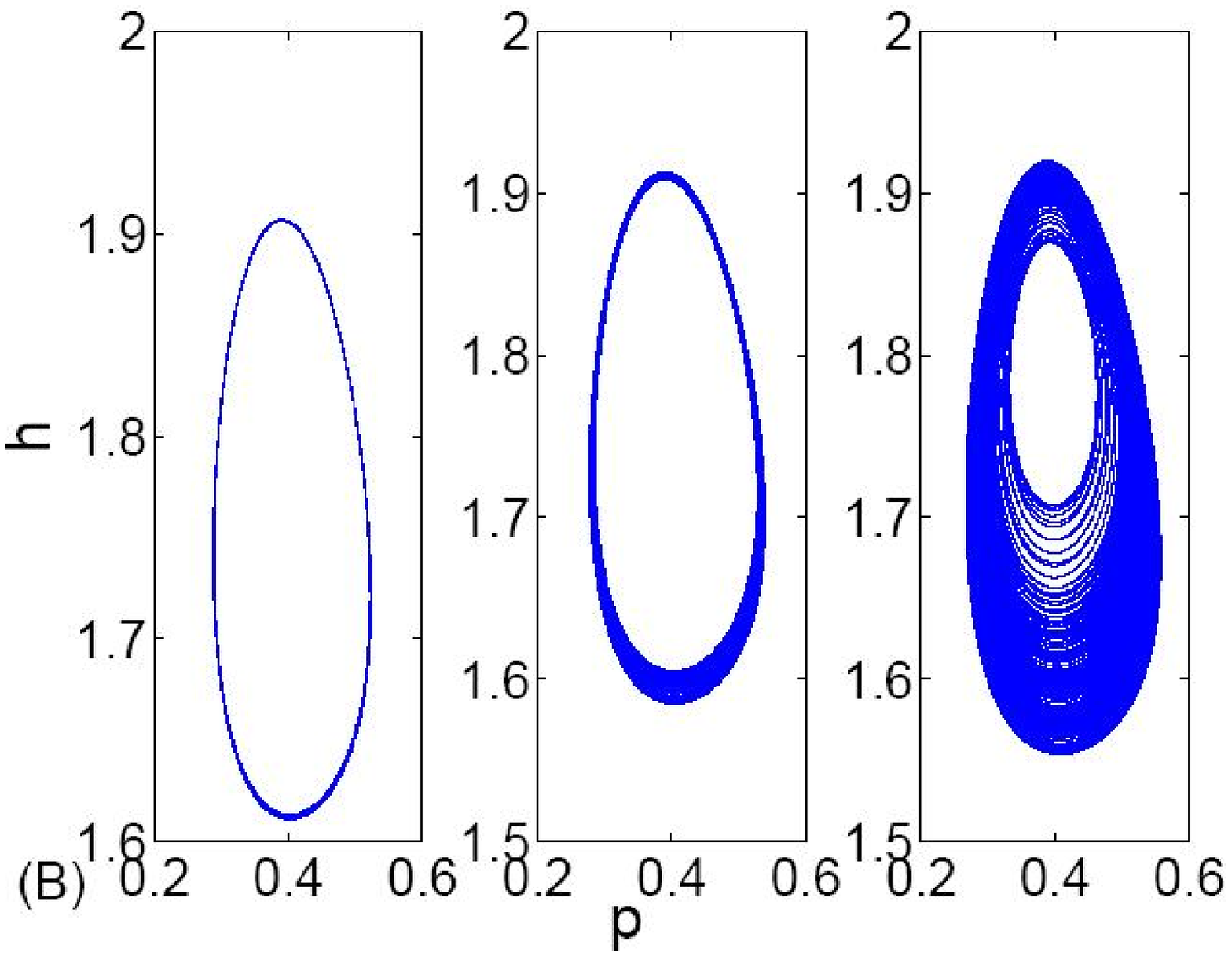}
\caption{\label{trajectories} The corresponding trajectories (from
left to right) for locations $(300,300)$, $(250,300)$, and
$(50,300)$ respectively. The parameters in (A), and (B) were the
same as these in Fig.~\ref{spiral}(A) and (B), respectively.}
\end{figure}

The corresponding trajectories of the spiral core and the spiral arm
(far away from the core center) at $y=300$ are shown in
Fig.~\ref{trajectories}, respectively. From Fig.~\ref{trajectories},
we can see that the spiral core is not completely fixed, but
oscillates with a large amplitude. However, as $d_h$ decreases to a
critical value, an unstable modulation develops in regions which is
far away from the spiral core (cf. the middle column of the
Fig.~\ref{trajectories}). These oscillations eventually grow large
enough to cause the spiral arm far away from the core to breakup
into complex multiple spiral waves, while the core region remains
stable (the corresponding movie can be viewed in the online
supplemental in Ref.~\cite{movie1} [partly $movie_{-}1$ and
$movie_{-}2$, and for $d_h=0.02$]). Figures~\ref{spiral}(B) and
\ref{trajectories}(B) show the dynamic behavior for $d_{h}=0.02$,
i.e., $\nu=0.4$. The regular trajectories far away from the core are
now the same as the region of the spatial chaos (cf. the middle
column of the Fig.~\ref{trajectories}). It is shown that an decrease in the 
diffusion ratio $\nu$ which leads to population oscillations of increasing amplitude (cf. the left column of the
Fig.~\ref{trajectories}). In the tradition explain that the minimum value of the population density decreases and population extinction becomes more probable due to stochastic environmental perturbations. However, from the spatial evolution of system~\eqref{eq:II2} (see Fig.~\ref{spiral}), the temporal variations of the density of different sub-population can become asynchronous and the events of local extinction can be compensated due to re-colonization (or diffusion) from other sites.

\begin{figure}[htp]
\includegraphics[width=7cm]{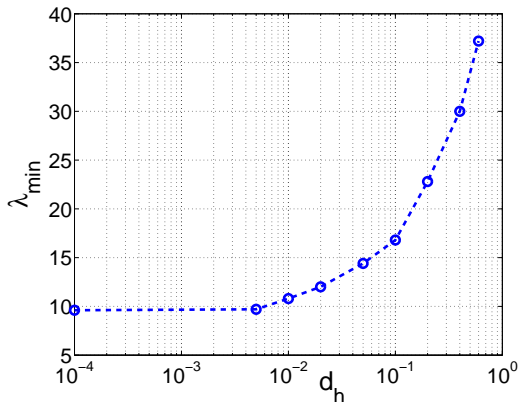}
\caption{\label{fig3} Dependence of the wavelength $\lambda_{min}$
 on the parameter $d_h$ for the system~\eqref{eq:II2} with $r=5.0$, $a=5.0$, $b=5.0$,
$m=0.6$, $d_{p}=0.05$, and $n=0.4$. Note the log scale for $d_h$.}
\end{figure}

\begin{figure}[h]
\includegraphics[width=7cm]{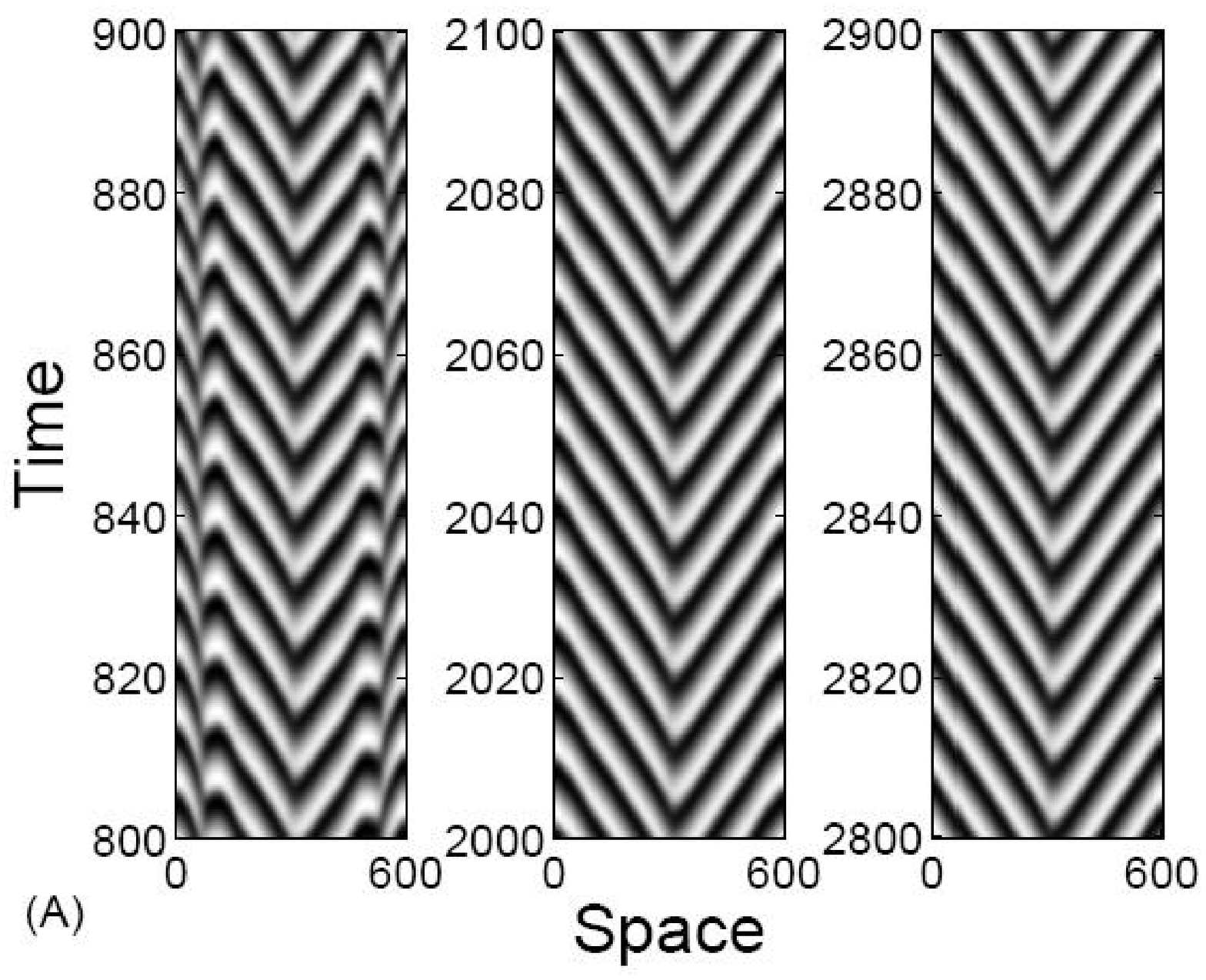}\\
\includegraphics[width=7cm]{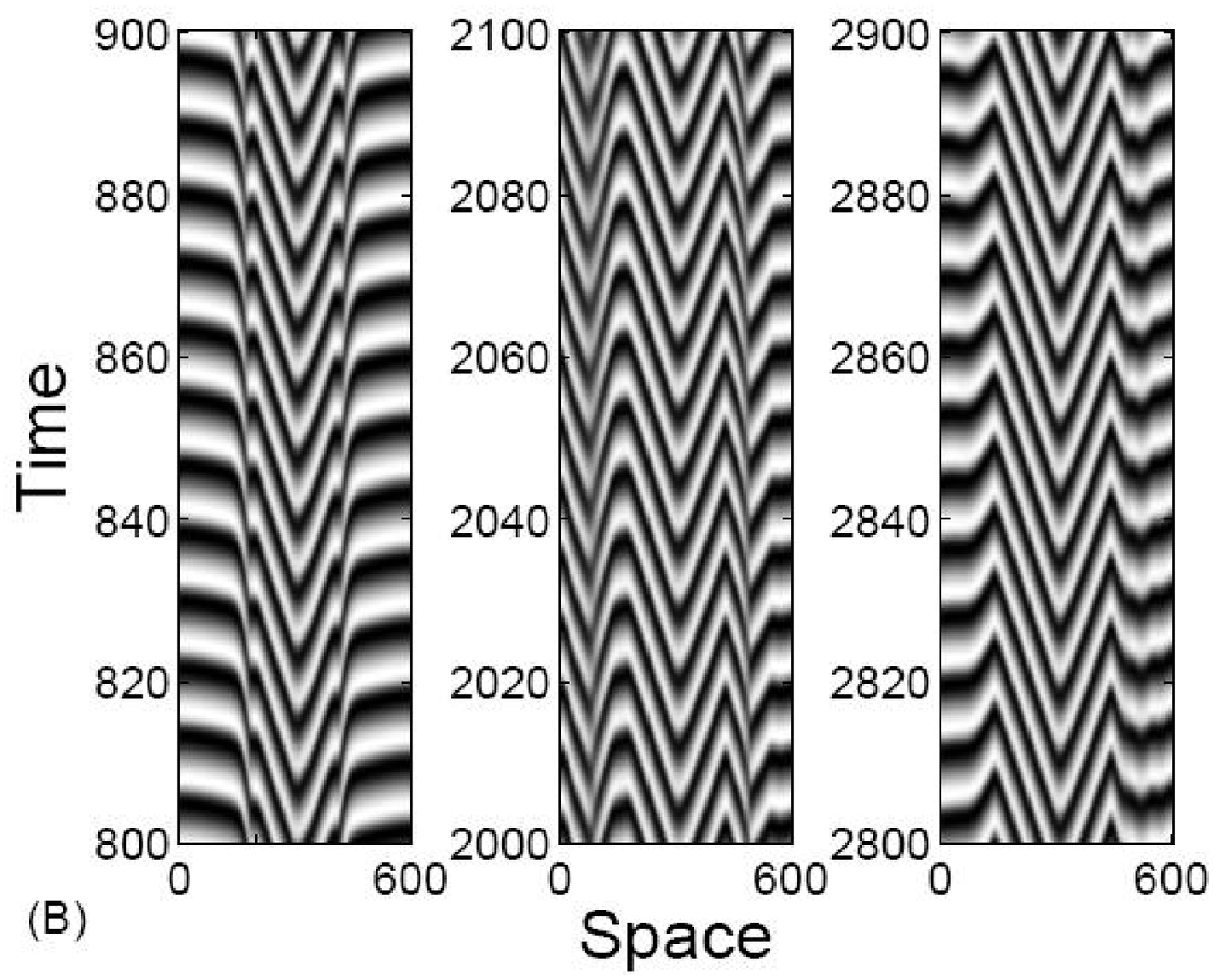}
\caption{\label{space-time} Space-time plots of variable $p$ for
different time and $d_h$. The parameters in (A), and (B) are the
same as those in Fig.~\ref{spiral}(A) and (B), respectively.}
\end{figure}

Furthermore, it is well known that the basic arguments in spiral
stability analysis can be carried out by reducing the system to one dimensional
space~\cite{Markus2004,PhysRevLett.82.1160,xie:026107,
PhysRevE.62.7708,PhysRevLett.80.4811}. Here we show some essential
properties of the spiral breakup resulting from the numerical
simulation. In the next section we will give the theoretical 
computation by using the eigenvalue spectra. In this model, it is worth noting that we do not neglect the oscillation of the dynamics in the core as shown in
Fig.~\ref{trajectories} due to the system exhibiting spatial periodic wave trains when the model is simulated in one-dimensional
space. Breakup occurs first far away from the core (the source of
waves). The spiral wave breaks towards the core until it gets to
some constant distance and then the surviving part of the spiral
wave stays stable. These minimal stable wavelengths are called
$\lambda_{min}$. So the one-parameter family may be described by a
dispersion curve $\lambda(d_h)$ (see Fig.~\ref{fig3}). The minimal
stable wavelength $\lambda_{min}$ of the spiral wave are shown in
Fig.~\ref{fig3} coming from the simulation in two dimensional space.
The results of Fig.~\ref{fig3} can be interpreted as follows: the
minimal stable wavelengths decrease with respect to the decrease of
$d_h$ but eventually stay at a relative constant value, which is
that the stable spiral waves are always existing for a larger region
values of $d_h$. Space-time plots at different times are shown in
Fig.~\ref{space-time} for two different $d_h$, i.e., different
$\nu$, which display the time evolution of the spiral wave along the
cross section in the two-dimensional images of Fig.~\ref{spiral}(A)
and (B). As shown in Fig.~\ref{space-time}(A) and (B) for $d_h=0.2$
and $d_h=0.02$ respectively, the waves far away from the core
display unstable modulated perturbation due to convective
instability~\cite{Markus2004,PhysRevLett.82.1160,xie:026107,
PhysRevE.62.7708,PhysRevLett.80.4811}, but this perturbation is
gradually advected to the left and right sides, and finally
disappears. The instability manifests itself to produce the wave
train breakup several waves from the far-field, as shown in
Figs.~\ref{space-time}(B).


\section{Spectra and Nonlinear bifurcation of the spiral wave}

In this section, we concentrate on the linear stability analysis of spiral wave by using the spectrum theory~\cite{PhysRevE.62.7708,sandstede:016217,Rademacher2007,2006Whell,SandstedephysicaD}.
From the results in Refs.~\cite{PhysRevE.62.7708,2006Whell} we know that the absolute spectrum must be computed numerically for any given reaction-diffusion systems. In practice, such computations only require discretization in one-dimensional space and compare with computing eigenvalues of the full stability problem on a large domain due to the spiral wave exhibitting traveling waves in the plane (see Fig.~\ref{space-time} about the space-time graphes). For spiral waves on the unbounded plane, the essential spectrum is also required to compute, since it determined only by the far-field wave trains of the spiral. The linear stability spectrum consists of point eigenvalues and the essential spectrum that is a continuous spectrum for spiral waves.

For sake of simplicity, the Eqs.~\eqref{subeq:II2a} and \eqref{subeq:II2b} can been written as following
\begin{subequations}
\label{eq:III3}
\begin{equation}
\frac{\partial p}{\partial t}=d_{p}\nabla^{2} p+g_{1}(p,h),\label{subeq:III2a}
\end{equation}
\begin{eqnarray}
\frac{\partial h}{\partial
t}=d_{h}\nabla^{2}
   h+g_{2}(p,h).\label{subeq:III2b}
\end{eqnarray}
\end{subequations}

Suppose that $(p^{*},h^{*})$ are a solutions and refer to them as steady spirals of Eq.~\eqref{eq:III3} that rotate rigidly with a constant angular velocity $\omega$, and that are asymptotically periodic along rays in the plane.  In a coratating coordinate frame, using the standardized analysis method for the spiral waves~\cite{SandstedephysicaD,2006Whell}, the Eq.~\eqref{eq:III3} is given by
\begin{subequations}
\label{eq:III4}
\begin{equation}
\frac{\partial p}{\partial t}=d_{p}\nabla^{2}_{\rho,\theta} p+\omega \frac{\partial p}{\partial \theta}+g_{1}(p^{*},h^{*}),\label{subeq:III4a}
\end{equation}
\begin{eqnarray}
\frac{\partial h}{\partial
t}=d_{h}\nabla^{2}_{\rho,\theta}
   h+\omega \frac{\partial h}{\partial \theta}+g_{2}(p^{*},h^{*}),\label{subeq:III4b}
\end{eqnarray}
\end{subequations}
where $(\rho,\theta)$ denote polar coordinates, spirals waves are relative equilibria, then the statianry solutions $p^{*}(\rho,\theta)$ and $h^{*}(\rho,\theta)$ both are $2\pi$-periodic functions with $\theta=\varphi-\omega t$. In Eqs.~\eqref{subeq:III4a} and \eqref{subeq:III4b} the operator $\nabla^{2}_{\rho,\theta}$ denotes $\partial_{\rho\rho}+\frac{1}{\rho} \partial_{\rho}+\frac{1}{\rho^2}\partial_{\theta\theta}$.

\subsection{Computation of spiral spectra}

Next, we commpute the leading part of its linear stability spectrum for the system~\eqref{eq:III4}. Consider the linearized evolution equation in the rotating frame, the eigenvalue problem of Eqs.~\eqref{subeq:III4a} and \eqref{subeq:III4b} associated with the planar spiral solutions $p^{*}(\rho,\theta)$ and $h^{*}(\rho,\theta)$ are given by
 \begin{subequations}
 \label{eq:III5}
 \begin{equation}
  d_{p}\nabla^{2}_{\rho,\theta}p+\omega \frac{\partial p}{\partial \theta} +g_{1}^{p}(p^{*},h^{*})p+ g_{1}^{h}(p^{*},h^{*})h=\lambda p,\label{subeq:III5a}
 \end{equation}
 \begin{eqnarray}
  d_{h}\nabla^{2}_{\rho,\theta}h+\omega \frac{\partial h}{\partial \theta} +g_{2}^{p}(p^{*},h^{*})p+ g_{2}^{h}(p^{*},h^{*})h=\lambda h,\label{subeq:III5a}
 \end{eqnarray}
 \end{subequations}
 where $g_{1}^{p}$, $\cdots$, $g_{2}^{h}$ denote the derivatives of
 the nonlinear functions and $ g_{1}^{p}(p,h)=r(1-p)-rp-\frac{ah}{1+bp}+\frac{abph}{(1+bp)^2}$, $g_{1}^{h}(p,h)=-\frac{ap}{1+bp}$, $g_{2}^{p}(p,h) =\frac {ah}{1+bp}-\frac {abph}{ \left( 1+bp \right) ^{2}}$,  and $g_{2}^{h}(p,h) =\frac {ap}{1+bp}-m-\frac {2fnh}{{n}^{2}+{h}^{2}}+\frac {2fn{h}^{3}}{ \left( {n}^{2}+{h}^{2}
  \right) ^{2}}$. We shall ignore isolated eigenvalues that belong to the point spectrum, instabilities caused by point eigenvalues lead to meanderingor drifting waves, or to an unstable tip motionin in excitable media and oscillation media~\cite{PhysRevE.62.7708,PhysRevLett.68.2090,PhysRevLett.72.164,PhysRevLett.72.2316}. This phenomenon is not shown in the present paper.
 Instead, we focus on the continuous spectrum that is responsible for the spiral wave breakup in the far field (see Fig.~\ref{spiral}(b)). By the results in Ref.~\cite{2006Whell}, it turns out that the boundary of the continuous spectrum depends only on the limiting equation for $\rho\rightarrow \infty$. Thus, we have that $\lambda$ is the boundary of the continuous spectrum if, and only if the limiting equation
 \begin{subequations}
 \label{eq:III21}
 \begin{equation}
  d_{p}\nabla^{2}_{\rho,\rho}p+\omega \frac{\partial p}{\partial \theta} +g_{1}^{p}(p^{*},h^{*})p+ g_{1}^{h}(p^{*},h^{*})h=\lambda p,\label{subeq:III5a}
 \end{equation}
 \begin{eqnarray}
  d_{h}\nabla^{2}_{\rho,\rho}h+\omega \frac{\partial h}{\partial \theta} +g_{2}^{p}(p^{*},h^{*})p+ g_{2}^{h}(p^{*},h^{*})h=\lambda h,\label{subeq:III5a}
 \end{eqnarray}
 \end{subequations}
 have solutions $p(\rho,\theta)$ and $h(\rho,\theta)$ for $(\rho,\theta)\in R^{+}\times[0,2\pi]$, which are bounded but does not 
 decay as $\rho\rightarrow \infty$. Since spiral waves are rotating waves in the plane, the wave train solutions have the form as $u(t,x,y)=u(\rho,\varphi-\omega t)$ for an appropriate wave numbers $k$ and temporal frequency $\omega$, where we assume that $u$ is $2\pi$-periodic in its argument so that $u(\xi)=u(\xi+2\pi)$ for all $\xi$ and $u=(p,h)^{\text{T}}$. Spiral waves converge to wave trains
 $u(\rho, \varphi -\omega t)\rightarrow u_{wt} (k\rho +\varphi-\omega t)$ for $\rho \rightarrow \infty$, which are corresponding to asymptotically Archimedean in the two-dimensional space. Assume that $k\neq 0$ and $\omega\neq 0$, and in this case, we can pass from the theoretical frame $\rho$ to the comoving frame $\xi=k\rho+\varphi -\omega t$ ($\xi\in \mathcal{R}$) in which the eigenvalue equation~\eqref{eq:III21} becomes
 \begin{subequations}
 \label{eq:III22}
 \begin{equation}
  d_{p}k^2 \nabla^{2}_{\xi,\xi}p+\omega p_{\xi} +g_{1}^{p}(u_{wt}(\xi))p+ g_{1}^{h}(u_{wt}(\xi))h=\lambda p,\label{subeq:III5a}
 \end{equation}
 \begin{eqnarray}
  d_{h}k^2\nabla^{2}_{\xi,\xi}h+\omega h_{\xi} +g_{2}^{p}(u_{wt}(\xi))p+ g_{2}^{h}(u_{wt}(\xi))h=\lambda h.\label{subeq:III5b}
 \end{eqnarray}
 \end{subequations}
 Indeed, any nontrivial solution $u(\xi)=(p(\xi), h(\xi))^{\text{T}}$ corresponding to the linearization eigenvalue problem~\eqref{eq:III22} give a solution $U(\rho,\cdot)$ of the eigenvalue problem for the temporal period map of \eqref{eq:III3} in the corotating frame via 
 \begin{equation}
 U(\rho,\cdot)=e^{\lambda t}u(k\rho-\omega t),  \qquad U(\rho,T)=e^{\lambda T}u(k\rho-2\pi). 
 \end{equation} 
We write the equations~\eqref{eq:III22} as the first-order systems
 \begin{eqnarray}\label{eq:23}
 \dfrac{dp}{d\rho}&=&p_{1}, \nonumber \\
  \dfrac{dh}{d\rho}&=&h_{1}, \nonumber \\
\dfrac{dp_1}{d\rho}&=&k^{-2}d_{p}^{-1}\big[\mu p-\omega p_{1}-g_{1}^{p}(u_{wt}(\xi))p- g_{1}^{h}(u_{wt}(\xi))h\big], \nonumber \\
 \dfrac{dh_1}{d\rho}&=&k^{-2}d_{h}^{-1}\big[\mu h-\omega h_{1}-g_{2}^{p}(u_{wt}(\xi))p- g_{2}^{h}(u_{wt}(\xi))h\big], \nonumber \\
 \end{eqnarray} 
 in the radial variable $\rho$.   Then the spatial
 eigenvalues or spatial Floquet exponents are deternined as the roots of the Wronskian
 \begin{equation}
  \mathcal{A}(\lambda, k):=\begin{pmatrix}
 0 & 0 & 1 & 0  \\
 0 & 0 & 0 & 1 \\
 \frac{1}{k^2 d_{p}}(\lambda-g_{1}^{p}(u_{wt}(\xi))) & -\frac{1}{k^2 d_{p}}g_{1}^{h}(u_{wt}(\xi))&  -\frac{1}{k^2 d_{p}}\omega & 0 \\
 -\frac{1}{k^2 d_{h}}g_{2}^{p}(u_{wt}(\xi)) & \frac{1}{k^2 d_{h}}(\lambda-g_{2}^{h}(u_{wt}(\xi))) & 0 & -\frac{1}{k^2 d_{h}}\omega
  \end{pmatrix},
 \end{equation}
 where $k\in \mathcal{R}$. The function $U(\rho,\cdot)=e^{\lambda t} e^{\text{i}k \rho}u_{0}(k\rho-\omega t)$ satisfies the equation~\eqref{eq:III3} when the spatial and temporal exponents $\text{i}k$ and $\lambda$ satisfy the complex dispersion relation $\text{det}(\mathcal{A}(\lambda,k)-\text{i}k)=0$ for $\lambda \in\mathcal{C}$. We call the $\text{i}k$ in spectrum of $A(\lambda, k)$ as spatial
 eigenvalues or spatial Floquet exponents.
 
 The stability of the spiral waves state $(p^{*},h^{*})$ on the plane is determined by the essential spectrum given by 
 \begin{equation}
 \Sigma_{ess}=\{\lambda \in\mathcal{C}; \text{det}(\mathcal{A}(\lambda,k)-\text{i}k)=0 \; \text{for some} \; k\in \mathcal{R}  \}.
 \end{equation} 

\begin{figure}[h]
(A)\includegraphics[width=5cm]{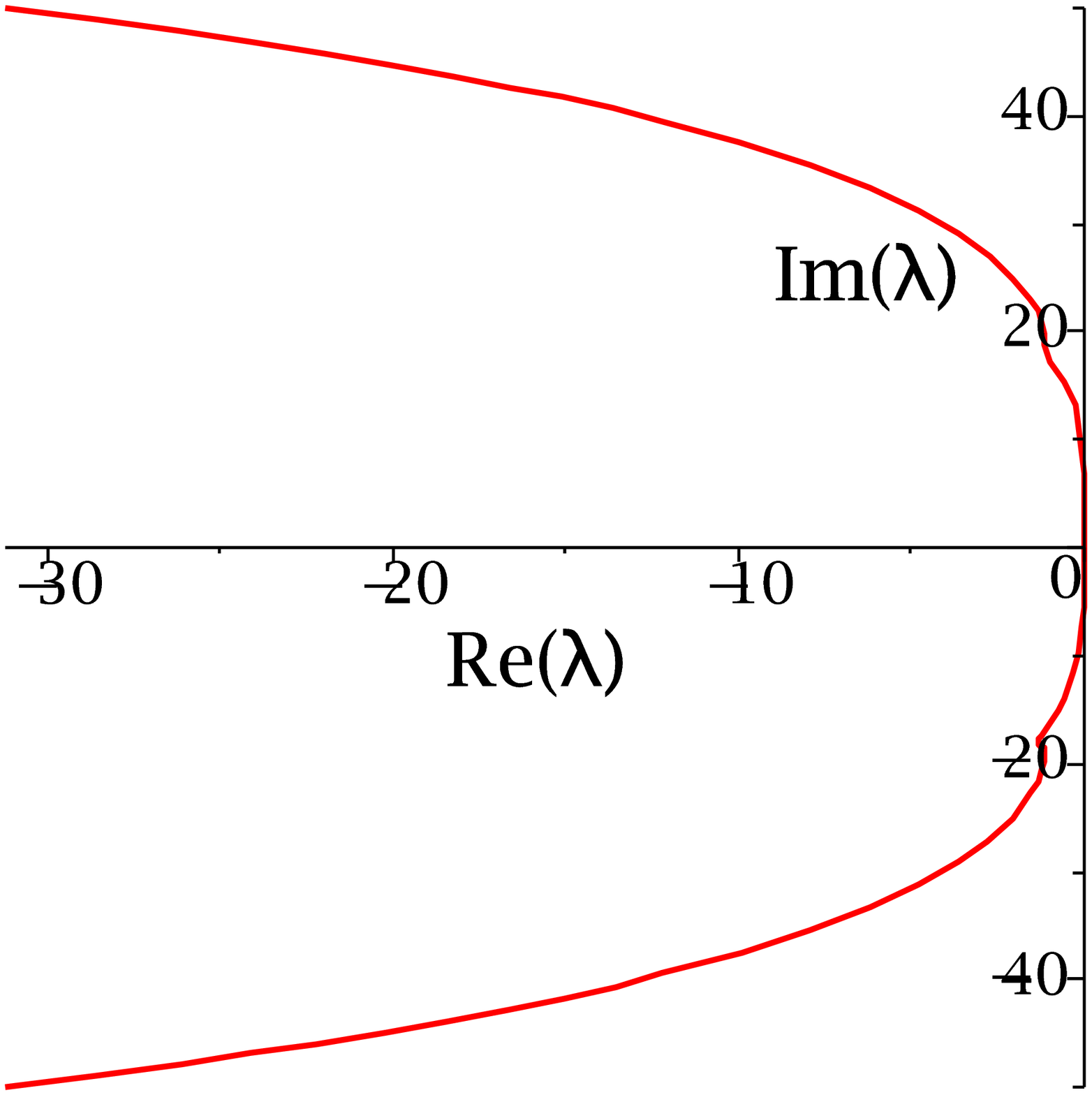}
(B)\includegraphics[width=5cm]{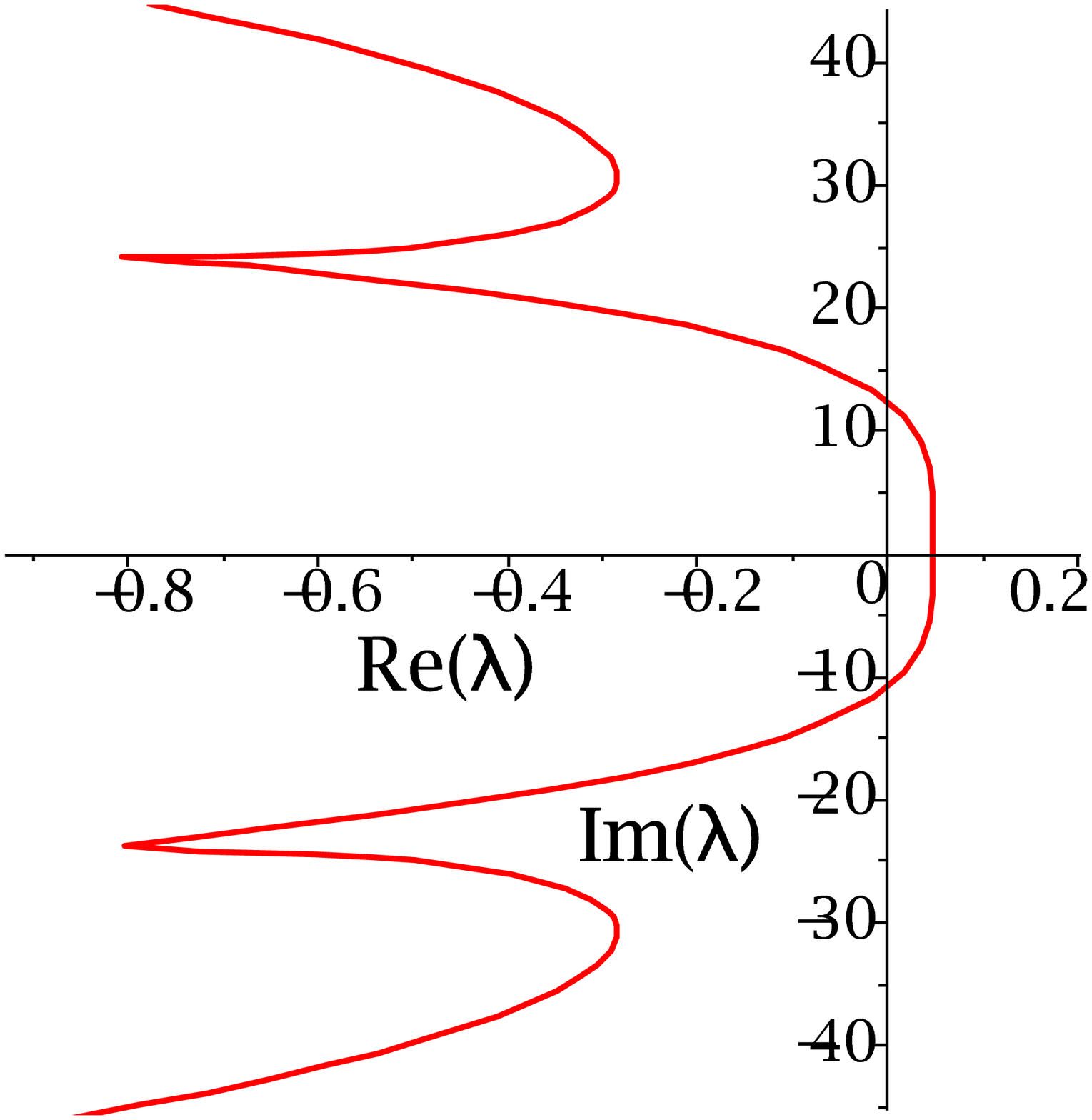}
\caption{\label{spectra} The essentail spectra of wave trains are obtained by  using the algorithms outlined in Refs.~\cite{Rademacher2007,sandstede:016217}. The parameters of (A) and (B) are corresponding to the values used in the simulations of Fig.~\ref{spiral}(A) and (B).}
\end{figure}

 Now, we compute the continuous spectrum with the equation~\eqref{eq:23} that are parameterized by the wave number $k$. For each $\lambda$, there are infinitely many stable and unstable spatial eigenvalues. We plot $\lambda$ in the complex plane  associated spatial spectrum, see Fig.~\ref{spectra}.
By the explaination of Sandstede \emph{et al}~\cite{sandstede:016217}, one would know that if the real part of the essentail spectra is positive, then the associated eigenmodes grow exponentially toward the boundary, i.e., they correspond to a far-field instability. Note that we find the essentail spectra are not sensitive to temporal frequency, $\omega$.

\subsection{Existence and properties of wave trains}\label{sub:1}

Suppose that a reaction-diffusion system on the one-dimensional space such that the variables equal to a homogeneous stationary solution. If the homogeneous steady-state destabilizes, then its linearization accommodates waves of the form $e^{\text{i}(kx-\omega t)}$ for certain values $k$ and $\omega$. Typically, near the transition to instability, small spatially periodic travelling waves arise for any wave number close to $k_{c}$, which is the critical wavenumber. Their wave speed is approximately equal to $\frac{\omega_{c}}{k_c}$, where $\omega_{c}$ is corresponding to $k_c$. In present paper, we focus exclusively on the situation where $\omega_{c}=0$ and $k_{c}\neq0$. The bifurcation with  $\omega_{c}=0$ and $k_{c}\neq0$ is known as the Turing bifurcation, and the bifurcating spatially periodic steady patterns are often referred to as Turing patterns. Another class of moved patterns will appear when the instabilities modulated by Hopf-Turing bifurcation, which is resemble a travelling waves.  Moreover, the common feature of the spiral waves in one-dimensional space mentioned above is the presence of wave trains which are spatially periodic travelling waves of the form $p_{wt}(kx-\omega t;k)$ and $h_{wt}(kx-\omega t;k)$, where $p_{wt}(\phi;k)$ and $h_{wt}(\phi;k)$ are $2\pi$-periodic about $\phi$. Typically, the spatial wavenumber $k$ and the temporal frequency $\omega$ are related via the nonlinear dispersion relation $\omega=\omega(k)$ so that the phase velocity is given by \begin{equation}
c_{p}=\frac{\omega}{k}.
\end{equation}
A second quantity related to the nonlinear dispersion relation is the group velocity, $c_{g}=\frac{d\omega}{dk}$, of the wave train which also play a central role in the spiral waves. The group velocity $c_{g}$ gives the speed of propagation of small localized wave-package perturbations of the wave train~\cite{sandstede:1}. Here, we are only concerned the existence of travelling wave solution. In fact, the spiral waves move at a constant speed outward from the core (see Fig.~\ref{space-time}), so that they have the mathematical form $p(x,t)=P(z)$, and $h(x,t)=H(z)$ where $z=x-c_{p}t$. Substituting these solution forms into Eq.~\eqref{eq:III3} gives the ODEs
\begin{subequations}
\label{eq:III6}
\begin{equation}
 d_{p}\frac{d^2 P}{dz^2}+c_{p}\frac{dP}{dz}+g_{1}(P,H)=0,\label{subeq:III6a}
\end{equation}
\begin{eqnarray}
 h_{p}\frac{d^2 H}{dz^2}+c_{p}\frac{dH}{dz}+g_{2}(P,H)=0.\label{subeq:III6a}
\end{eqnarray}
\end{subequations}

Here, we investigate numerically the existence, speed and wavelength of travelling wave patterns. Our approach is to use the bifurcation package Matcont 2.4~\cite{Matcont} to study the pattern ODEs~\eqref{eq:III6}. To do this, the most natural bifurcation parameters are the wave speed $c_{p}$ and $f$, but they give no information about the stability of travelling wave as solutions of the model PDEs~\eqref{eq:III3}.

\begin{figure}[h]
\includegraphics[width=7.2cm]{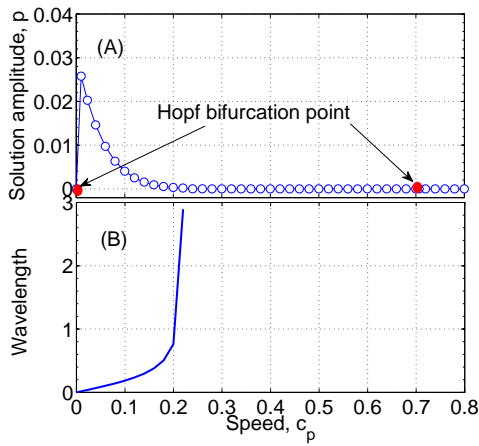}
\caption{\label{figspeed} Typical bifurcation diagrams for the pattern ODEs~\eqref{eq:III6}. (A) The spatially periodic travelling waves of system~\eqref{eq:III3} is existence. The changes in stability occur via Hopf bifurcation, from which a branch of periodic orbits emanate. Thus  unstable travelling waves appear. (B) Maximum stable wavelength along the bifurcation parameter$c_{p}$, i.e., $k\neq0, \omega \neq 0$. The parameter values in (A) and (B) are the same as Fig.~\ref{spiral}(A).}
\end{figure}

\begin{figure}[h]
(A)\includegraphics[width=7.2cm]{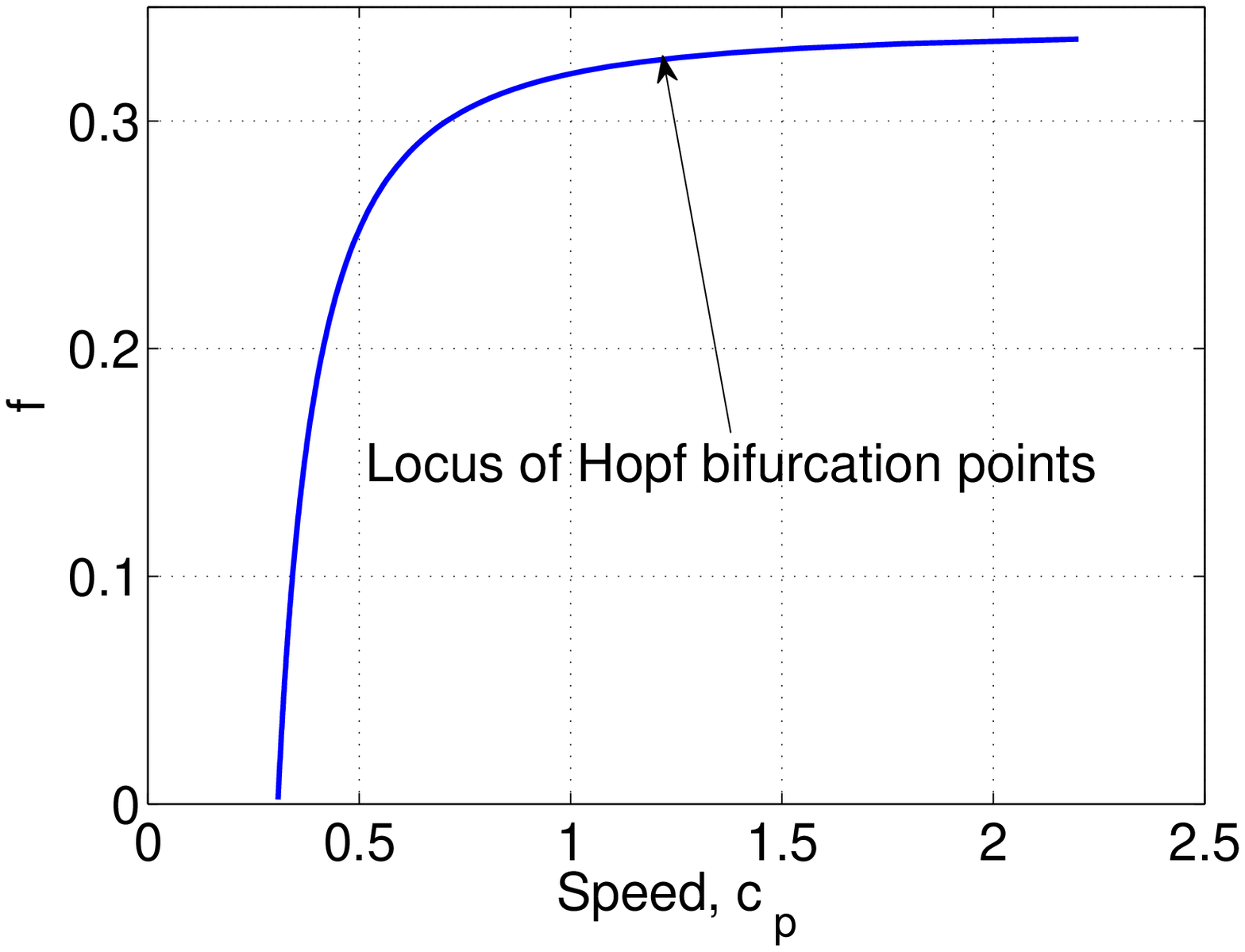}
(B)\includegraphics[width=7.2cm]{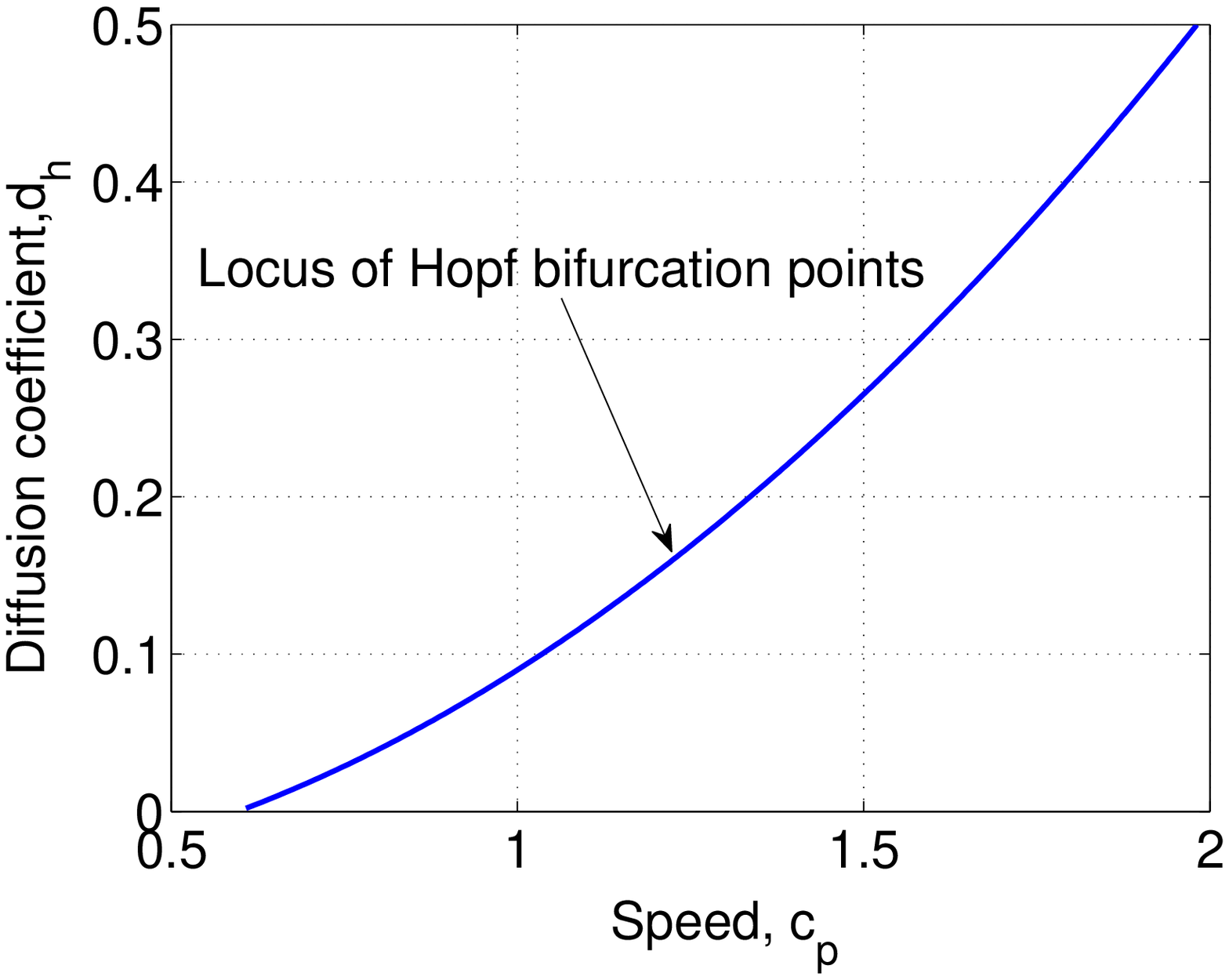}
\caption{\label{parameterspace} An illustration of the variations in parameter space of the pattern ODEs~\eqref{eq:III6}. We plot the loci of Hopf bifurcation points. (A) $f-c_{p}$ planes; (B) $d_{h}-c_{p}$ planes. The parameter values in (A) and (B) are the same as Fig.~\ref{spiral}(A).}
\end{figure}

Our starting point is the homogeneous steady state of Eq.~\eqref{eq:III6} with in the domain III of Fig.~\ref{Bifurcationdiagram}. The typical bifurcation diagrams are illustrated in Fig.~\ref{figspeed}, which shows that steady spatially peroidic travelling waves  exist for the larger values of the speed $c_p$, but it is unstable for small values of $c_p$. The changes in stability occur via Hopf bifurcation, from which a branch of periodic orbits emanate. Note that here we use the terms ``stable'' and ``unstable'' as referring to the ODEs system~\eqref{eq:III6} rather than the model PDEs. Fig.~\ref{figspeed}(B) illustrates the maximun stable wavelength against the bifucation parameter, speed $c_p$, and the small amplitudes have very long wavelength. It is known that $c_{p}=\frac{\omega}{k}$, hence the tavelling wave solution  exist when the $c_{p}\neq 0$, i.e., $k\neq0, \omega \neq 0$. Using Matcont 2.4 package, it is possible to track the locus of the Hopf bifurcation points and the Limit point (fold) bifurcation in a parameter plane, and a typical example of this for the $c_{p}$-$f$ and $c_{p}$-$d_{h}$ plane are illustrated in Fig.~\ref{parameterspace}. The travelling wave solutions exist for values of $c_p$ and $f$ lying in left of Hopf bifurcation locus (see Fig.~\ref{parameterspace}(A)). The same structure about the $c_p$-$d_h$ plane is shown in Fig.~\ref{parameterspace}(B). These reuslts confirm our previous analysis coming from the algebra computation (see Fig.~\ref{Bifurcationdiagram}) and the numerical results~(see Fig.~\ref{space-time}).


\section{conclusions and discussion}\label{conlusions}

We have investigated a spatially extended plankton ecological system within
 two-dimensional space and found that its spatial
patterns exhibit spiral waves dynamics and spatial chaos patterns.
Specially, the scenario of the spatiotemporal chaos patterns arising from the
far-field breakup is observed. Our research is based on numerical
analysis of a kinematic mimicking the diffusion in the dynamics of
marine organisms, coupled to a two component plankton model on the 
level of the community. By
increasing (decreasing) the diffusion ratio of the two variables, the spiral arm first broke up into a
turbulence-like state far away from the core center, but which do not
invade the whole space.  From the previous studies in the Belousov-Zhabotinsky reaction, we know the reason causing this phenomenon can be
illuminated theoretically by the M. B\"{a}r and L. Brusch~\cite{Markus2004,PhysRevLett.82.1160}, as well as by using the spectrum theory that poses by B. Sandstede, A. Scheel \emph{et al}~\cite{PhysRevE.62.7708,sandstede:016217,Rademacher2007,Wheeler}. The far-field breakup
can  be verified in field observation and is useful to understand
the population dynamics of oceanic ecological systems. Such as that
under certain conditions the interplay between wake (or ocean)
structures and biological growth leads to plankton blooms inside
mesoscale hydrodynamic vortices that act as incubators of primary
production. From Fig.~\ref{spiral} and corresponding the movies, we see that spatial peridic bloom appear in the phytoplankton populations, and the details of spatial evolution of the distribution of the phytoplankton population during one bloom cycle, respectively.

In Ref.~\cite{MaartenBoerlijst}, the authors study the optimal control of the model~\eqref{eq:II2} from the spatiotemporal chaos to spiral waves by the parameters for fish predation treated as a multiplicative control variable. Spatial order  emerges in a range of spatial models of multispecies
interactions. Unsurprisingly, spatial models of multispecies systems often
manifests very different behaviors from their mean-field counterparts. 
Two important general features of spatial models of multispecies systems are
that they allow the possibility of global persistence in spite of local extinctions and 
so are usually more stable than their mean-field equivalents, and have a tendency to self-organzie spatially or regular spatiotemporal patterns~\cite{Dieckmann,MaartenBoerlijst}. The spatial structures
produces nonrandom spatial patterns such as spiral waves and spatiotemporal chaos at scales much larger than the scale of interaction among individuals level. These structures are not explicitly coded but emerge from local interaction among individuals and local diffusion. 

 As we know
that plankton plays an important role in the marine ecosystem and
the climate, because of their participation in the global carbon and
nitrogen cycle at the base of the food chain~\cite{Duinker1994}.
From the review~\cite{Pascual2005}, a recently developed ecosystem
model incorporates different phytoplankton functional groups and
their competition for light and multiple nutrients. Simulations of
these models at specific sites to explore future scenarios suggest
that global environmental change, including global-warming-induced
changes, will alter phytoplankton community structure and hence
alter global biogeochemical cycles~\cite{Litchman}. The coupling of
spatial ecosystem model to global climate raises again a series of
open questions on the complexity of model and relevant spatial
scales. So the study of spatial model with large-scale is more
important in the ecological system. Basing on numerical simulation
on the spatial model, we can draft that the oceanic ecological
systems show permanent spiral waves and spatiotemporal chaos in large-scale
over a range of parameter values $d_h$, which indicates that
periodically sustained plankton blooms in the local area. As with all areas of evolutionary biology, theoretical development advances more quickly than does empiraical evidence. The most powerful empirical approach is to conduct experiments in which the spatial pattern can be measured directly, but this is difficulties in the design. However, we can indirectly measured these phenomenona by the simulation and compared with the satellite pictures. For example, the
spatiotemporal chaos patterns agree with the perspective observation of
the Fig.~3 in Ref.~\cite{Pascual2005}. Also, some satellite imageries [\href{http://oceancolor.gsfc.nasa.gov}{http://oceancolor.gsfc.nasa.gov}]
have displayed spiral patterns that
represent the phytoplankton [the chlorophyll] biomass and thus
demonstrated that plankton patterns in the ocean occur on much
broader scales and therefore mechanisms thought diffusion should be
considered.

\begin{acknowledgments}
This work is supported by the National
Natural Science Foundation of China under Grant No. 10471040 and the
Natural Science Foundation of Shan'xi Province Grant No. 2006011009.
\end{acknowledgments}

\newpage 
\bibliographystyle{unsrt} 

\begin{thebibliography}{10}

\bibitem{amritkar:258102}
R.~E. Amritkar and Govindan Rangarajan.
\newblock Spatially synchronous extinction of species under external forcing.
\newblock {\em Phys. Rev. Lett.}, 96(25):258102, 2006.

\bibitem{pekalski:021913}
Andrzej Pekalski and Michel Droz.
\newblock Self-organized packs selection in predator-prey ecosystems.
\newblock {\em Phys. Rev. E}, 73(2):021913, 2006.

\bibitem{Sayama2003}
Y.-Y.~H. Sayama, M.~A.~M. de~Aguiar, and M.~Baranger.
\newblock Interplay between turing pattern formation and domain coarsening in
  spatially extended population models.
\newblock {\em FORMA}, 18:19, 2003.

\bibitem{gilad:098105}
E.~Gilad, J.~von Hardenberg, A.~Provenzale, M.~Shachak, and E.~Meron.
\newblock Ecosystem engineers: From pattern formation to habitat creation.
\newblock {\em Phys. Rev. Lett.}, 93(9):098105, 2004.

\bibitem{mobilia2006}
Mark~J Washenberger, Mauro Mobilia, and Uwe~C T\"{a}uber.
\newblock Influence of local carrying capacity restrictions on stochastic
  predator\&ndash;prey models.
\newblock {\em J. Phys.: Cond. Matt.}, 19(6), 2007.

\bibitem{mobilia2005}
Mauro Mobilia, Ivan Georgiev, and Uwe T\"{a}uber.
\newblock Phase transitions and spatio-temporal fluctuations in stochastic
  lattice lotka�cvolterra models.
\newblock {\em J. Stat. Phys.}, 128(1):447--483, 2007.

\bibitem{Blasius1999}
Bernd Blasius, Amit Huppert, and Lewi Stone.
\newblock Complex dynamics and phase synchronization in spatially extended
  ecological systems.
\newblock {\em Nature}, 399(6734):354--359, 1999.

\bibitem{PhysRevLett.87.198101}
J.~von Hardenberg, E.~Meron, M.~Shachak, and Y.~Zarmi.
\newblock Diversity of vegetation patterns and desertification.
\newblock {\em Phys. Rev. Lett.}, 87(19):198101, 2001.

\bibitem{PhysRevE.67.056602}
A.~Provata and G.~A. Tsekouras.
\newblock Spontaneous formation of dynamical patterns with fractal fronts in
  the cyclic lattice lotka-volterra model.
\newblock {\em Phys. Rev. E}, 67(5):056602, 2003.

\bibitem{PhysRevE.64.021915}
Alexander~B. Medvinsky, Irene~A. Tikhonova, Rubin~R. Aliev, Bai-Lian Li,
  Zhen-Shan Lin, and Horst Malchow.
\newblock Patchy environment as a factor of complex plankton dynamics.
\newblock {\em Phys. Rev. E}, 64(2):021915, 2001.

\bibitem{medvinsky:311}
Alexander~B. Medvinsky, Sergei~V. Petrovskii, Irene~A. Tikhonova, Horst
  Malchow, and Bai-Lian Li.
\newblock Spatiotemporal complexity of plankton and fish dynamics.
\newblock {\em SIAM Review}, 44:311--370, 2002.

\bibitem{Gurney1998}
W.~S.~C. Gurney, A.~R. Veitch, I.~Cruickshank, and G.~McGeachin.
\newblock circles and spirals: population persistence in a spatially explicit
  predator�cprey model.
\newblock {\em Ecology}, 79(7):2516--2530, 1998.

\bibitem{Murray2002}
J.~D. Murray.
\newblock {\em Mathematical biology}.
\newblock Interdisciplinary applied mathematics. Springer, New York, 3rd
  edition, 2002.

\bibitem{Petrovskii}
Sergei Petrovskii, Bai-Lian Li, and Horst Malchow.
\newblock Transition to spatiotemporal chaos can resolve the paradox of
  enrichment.
\newblock {\em Ecological Complexity}, 1:37--47, 2004.

\bibitem{Laurence}
Laurence Armi, Dave Hebert, Neil Oakey, James Price, Philip~L. Richardson,
  Thomas Rossby, and Barry Ruddick.
\newblock The history and decay of a mediterranean salt lens.
\newblock {\em Nature}, 333(6174):649--651, 1988.

\bibitem{EsaRanta11281997}
Esa Ranta, Veijo Kaitala, and Per Lundberg.
\newblock {The Spatial Dimension in Population Fluctuations}.
\newblock {\em Science}, 278(5343):1621--1623, 1997.

\bibitem{Luckinbill}
L~S Luckinbill.
\newblock The effects of space and enrichment on a predator-prey system.
\newblock {\em Ecology}, 55:1142--1147, 1974.

\bibitem{Holyoak}
M~Holyoak.
\newblock Effects of nurient enrichment on predator-prey metapopulation
  dynamics.
\newblock {\em J Anim. Ecol.}, 69:985--997, 2000.

\bibitem{Jansen}
Vincent~A.A. Jansen.
\newblock Regulation of predator-prey systems through spatial interactions: a
  possible solution to the paradox of enrichment.
\newblock {\em Oikos}, 74:384–390, 1995.

\bibitem{Jansen2000}
Vincent~A.A. Jansen and Alun~L. Lloyd.
\newblock Local stability analysis of spatially homogeneous solutions of
  multi-patch systems.
\newblock {\em J. Math. Biol.}, 41:232–252, 2000.

\bibitem{Jansen2001}
Vincent~A.A. Jansen.
\newblock The dynamics of two diffusively coupled predator–prey populations.
\newblock {\em Theor. Popul. Biol.}, 59:119–131, 2001.

\bibitem{Allen2001}
J.~C. Allen, W.~M. Schaffer†, and D.~Rosko.
\newblock Chaos reduces species extinction by amplifying local population
  noise.
\newblock {\em Nature}, 364:229–232, 1993.

\bibitem{RevModPhys.65.851}
M.~C. Cross and P.~C. Hohenberg.
\newblock Pattern formation outside of equilibrium.
\newblock {\em Rev. Mod. Phys.}, 65(3):851, 1993.

\bibitem{PhysRevLett.87.068101}
Kyoung~J. Lee, Raymond~E. Goldstein, and Edward~C. Cox.
\newblock Resetting wave forms in dictyostelium territories.
\newblock {\em Phys. Rev. Lett.}, 87(6):068101, 2001.

\bibitem{Sawai2005}
Satoshi Sawai, Peter~A. Thomason, and Edward~C. Cox.
\newblock An autoregulatory circuit for long-range self-organization in
  dictyostelium cell populations.
\newblock {\em Nature}, 433(7023):323--326, 2005.

\bibitem{winfree:303}
Arthur~T. Winfree.
\newblock Varieties of spiral wave behavior: An experimentalist's approach to
  the theory of excitable media.
\newblock {\em Chaos}, 1(3):303--334, 1991.

\bibitem{Biktashev}
V.~N. Biktashev, J.~Brindley, A.~V. Holden, and M.~A. Tsyganov.
\newblock Pursuit-evasion predator-prey waves in two spatial dimensions.
\newblock {\em Chaos}, 14(4):988--994, 2004.

\bibitem{Garvie2007}
M~Garvie.
\newblock Finite-difference schemes for reaction-diffusion equation modeling
  predato-prey interactions in matlab.
\newblock {\em Bull. Math. Biol.}, 69:931--956, 2007.

\bibitem{garvie:775}
Marcus~R. Garvie and Catalin Trenchea.
\newblock Optimal control of a nutrient-phytoplankton-zooplankton-fish system.
\newblock {\em SIAM J. Contr. and Opti.}, 46(3):775--791, 2007.

\bibitem{Markus2004}
Markus B\"{a}r and Lutz Brusch.
\newblock Breakup of spiral waves caused by radial dynamics: Eckhaus and finite
  wavenumber instabilities.
\newblock {\em New Journal of Physics}, 6:5, 2004.

\bibitem{PhysRevLett.82.1160}
Markus B\"ar and Michal Or-Guil.
\newblock Alternative scenarios of spiral breakup in a reaction-diffusion model
  with excitable and oscillatory dynamics.
\newblock {\em Phys. Rev. Lett.}, 82(6):1160--1163, 1999.

\bibitem{Craigjohnson2002}
Craig~R Johnson and Maarten~C Boerlijst.
\newblock Selection at the level of the community: the importance of spatial
  structure.
\newblock {\em Trends Ecol. and Evol.}, 17:83--90, 2002.

\bibitem{Pascual1993}
M~Pascual.
\newblock Diffusion-induced chaos in a spatial predator–prey system.
\newblock {\em Proc. R. Soc. Lond. B}, 251:1–7, 1993.

\bibitem{Sherratt1997}
J.~A. Sherratt, B.~T. Eagan, and M.~A. Lewis.
\newblock Oscillations and chaos behind predator–prey invasion: mathematical
  artifact or ecological reality?
\newblock {\em Phil. Trans. R. Soc. Lond. B}, 352:21–38, 1997.

\bibitem{Petrovskii2001}
S.~V. Petrovskii and H.~Malchow.
\newblock Wave of chaos: new mechanism of pattern formation in spatio-temporal
  population dynamics.
\newblock {\em Theor. Popul. Biol.}, 59:157–174, 2001.

\bibitem{Sherratt2001}
J.~A. Sherratt.
\newblock Periodic travelling waves in cyclic predator–prey systems.
\newblock {\em Ecol. Lett.}, 4:30–37, 2001.

\bibitem{Savill1998}
N.~J. Savill and P.~Hogeweg.
\newblock Spatially induced speciation prevents extinction: the evolution of
  dispersal distance in oscillatory predator-prey models.
\newblock {\em Proc. R. Soc. Lond. B}, 265(1390):25--32, 1998.

\bibitem{Abraham1998}
E.~R. Abraham.
\newblock The generation of plankton patchiness by turbulent stirring.
\newblock {\em Nature}, 391:577--580, 1998.

\bibitem{Folt1999}
Carol~L. Folt and Carolyn~W. Burns.
\newblock Biological drivers of zooplankton patchiness.
\newblock {\em Nature}, 14:300--305, 1999.

\bibitem{MartenScheffer01011991}
M~Scheffer.
\newblock Should we expect strange attractors behind plankton dynamics and if
  so, should we bother?
\newblock {\em J. Plankton Res.}, 13:1291--1305, 1991.

\bibitem{Hanski1993}
I.~Hanski, P.~Turchin, E.~Korplmakl, and H.~Henttonen.
\newblock Population oscillations of boreal rodents: regulation by mustelid
  predators leads to chaos.
\newblock {\em Nature}, 364:232–235, 1993.

\bibitem{Ellner1995}
S.~Ellner and P.~Turchin.
\newblock Chaos in a noisy world: new methods and evidence from time-series
  analysis.
\newblock {\em Am. Nat.}, 145:343–375, 1995.

\bibitem{Dennis2001}
B.~Dennis, R.~A. Desharnais, J.~M. Cushing, S.~M. Henson, and R.~F. Costantino.
\newblock Estimating chaos and complex dynamics in an insect population.
\newblock {\em Ecol. Monogr.}, 71:277–303, 2001.

\bibitem{Scheffer1991a}
M~Scheffer.
\newblock Fish and nutrients interplay determines algal biomass: A minimal
  model.
\newblock {\em Oikos}, 62:271--282, 1991.

\bibitem{Malchow1993}
H.~Malchow.
\newblock Spatio-temporal pattern formation in nonlinear non-equilibrium
  plankton dynamics.
\newblock {\em Procc. R. Soc. Lond. B}, 251:103, 1993.

\bibitem{MercedesPascual}
M.~Pascual.
\newblock Diffusion-induced chaos in a spatial predator-prey system.
\newblock {\em Procc. R. Soc. Lond. B}, 251:1--7, 1993.

\bibitem{Scheel2003}
Arnd Scheel.
\newblock Radialy symmetric patterns of reaction-diffusion systems.
\newblock {\em Mem. Amer. Math. Soc.}, 165:86, 2003.

\bibitem{Satnoianu}
R~A Satnoianu, M~Menzinger, and P~K Maini.
\newblock Turing instabilities in general system.
\newblock {\em J. Math. Biol.}, 41:493--512, 2000.

\bibitem{liuqx2007}
Quan-Xing Liu, Bai-Lian Li, and Zhen Jin.
\newblock Resonant patterns and frequency-locked induced by additive noise and
  periodically forced in phytoplankton-zooplankton system, 2007.

\bibitem{Jorgensen}
Sven~Erik J{\o}rgensen and G.~Bendoricchio.
\newblock {\em Fundamentals of ecological modelling}.
\newblock Developments in environmental modelling; 21. Elsevier, Amsterdam; New
  York, 3rd edition, 2001.

\bibitem{Okubo1980}
Akira Okubo.
\newblock {\em Diffusion and ecological problems: mathematical models}.
\newblock Biomathematics; v. 10. Springer-Verlag, Berlin; New York, 1980.

\bibitem{Sugihara}
George Sugihara and Robert~M. May.
\newblock Nonlinear forecasting as a way of distinguishing chaos from
  measurement error in time series.
\newblock {\em Nature}, 344(6268):734--741, 1990.

\bibitem{Turing1952}
A.~M. Turing.
\newblock The chemical basis of morphogenesis.
\newblock {\em Philosophical Transactions of the Royal Society of London.
  Series B, Biological Sciences}, 237(641):37--72, 1952.

\bibitem{movie1}


\bibitem{xie:026107}
Fagen Xie, Dongzhu Xie, and James~N. Weiss.
\newblock Inwardly rotating spiral wave breakup in oscillatory
  reaction-diffusion media.
\newblock {\em Phys. Rev. E}, 74(2):026107, 2006.

\bibitem{PhysRevE.62.7708}
Bj\"orn Sandstede and Arnd Scheel.
\newblock Absolute versus convective instability of spiral waves.
\newblock {\em Phys. Rev. E}, 62(6):7708--7714, 2000.

\bibitem{PhysRevLett.80.4811}
S.~M. Tobias and E.~Knobloch.
\newblock Breakup of spiral waves into chemical turbulence.
\newblock {\em Phys. Rev. Lett.}, 80(21):4811--4814, 1998.

\bibitem{Ouyang1996}
Q.~Ouyang and J.~M. Flesselles.
\newblock Transition from spirals to defect turbulence driven by a convective
  instability.
\newblock {\em Nature}, 379(6561):143--146, 1996.

\bibitem{Ouyang2000}
Qi~Ouyang, H.~L. Swinney, and G.~Li.
\newblock Transition from spirals to defect-mediated turbulence driven by a
  doppler instability.
\newblock {\em Phys. Rev. Lett.}, 84(5):1047--1050, 2000.

\bibitem{sandstede:016217}
Bj\"{o}rn Sandstede and Arnd Scheel.
\newblock Curvature effects on spiral spectra: Generation of point eigenvalues
  near branch points.
\newblock {\em Phys. Rev. E}, 73:016217, 2006.

\bibitem{Rademacher2007}
Jens~D.M. Rademacher, Bj\"{o}rn Dandstede, and Arnd Scheel.
\newblock Computing absolute and essential spectra using continuation.
\newblock {\em Physics D}, 229:166--183, 2007.

\bibitem{2006Whell}
P.~{Wheeler} and D.~{Barkley}.
\newblock {Computation of Spiral Spectra}.
\newblock {\em SIAM J Appl. Dynam. Syst.}, 2006.

\bibitem{SandstedephysicaD}
Bj\"{o}rn Sandstede and Arnd Scheel.
\newblock Absolute and convective instabilities of waves on unbounded and large
  bound domians.
\newblock {\em Physics D}, 145:233--277, 2000.

\bibitem{PhysRevLett.68.2090}
Dwight Barkley.
\newblock Linear stability analysis of rotating spiral waves in excitable
  media.
\newblock {\em Phys. Rev. Lett.}, 68(13):2090--2093, 1992.

\bibitem{PhysRevLett.72.164}
Dwight Barkley.
\newblock Euclidean symmetry and the dynamics of rotating spiral waves.
\newblock {\em Phys. Rev. Lett.}, 72(1):164--167, 1994.

\bibitem{PhysRevLett.72.2316}
Igor Aranson, Lorenz Kramer, and Andreas Weber.
\newblock Core instability and spatiotemporal intermittency of spiral waves in
  oscillatory media.
\newblock {\em Phys. Rev. Lett.}, 72(15):2316--2319, 1994.

\bibitem{sandstede:1}
Bj\"{o}rn Sandstede and Arnd Scheel.
\newblock Defects in oscillatory media: Toward a classification.
\newblock {\em SIAM J. Appl. Dynam. Syst.}, 3(1):1--68, 2004.

\bibitem{Matcont}
A~Dhooge, W~Govaerts, Yu~A Kuznetsov, W~Mestrom, A~M Riet, and B~Sautois.
\newblock {\em Matcont and Cl-Matcont: Continuation toolboxes in Matlab}.
\newblock Utrecht University, The Netherlands, 2006.

\bibitem{Wheeler}
Paul Wheeler and Dwight Barkley.
\newblock Computation of spiral spectra.
\newblock {\em SIAM J. Appl. Dynam. Syst.}, 5:157--177, 2006.

\bibitem{MaartenBoerlijst}
Maarten~C. Boerlijst.
\newblock {\em The Geometry of Ecological Interactions: Simplifying Spatial
  Complexity}, chapter Spirals and spots: Novel Evolutionary Phenomena through
  spatial self-structuring, pages 171--182.
\newblock Cambridge University Press, 2000.

\bibitem{Dieckmann}
Ulf Dieckmann, Richard Law, and Johan A~J Metz.
\newblock {\em The Geometry of Ecological Interactions: Simplifying Spatial
  Complexity}.
\newblock Cambridge University Press, 2000.

\bibitem{Duinker1994}
J.~Duinker and G.~Wefer.
\newblock Das co2-problem und die rolle des ozeans.
\newblock {\em Naturwissenschaften}, 81(6):237--242, 1994.

\bibitem{Pascual2005}
M.~Pascual.
\newblock Computational ecology: From the complex to the simple and back.
\newblock {\em Plos Comput. Biol.}, 1(2):101--105, 2005.

\bibitem{Litchman}
E.~Litchman, C.~A. Klausmeier, J.~R. Miller, O.~M. Schofield, and P.~G.
  Falkowski.
\newblock Multi-nutrient, multi-group model of present and future oceanic
  phytoplankton communities.
\newblock {\em Biogeosciences}, 3(4):585--606, 2006.

\end{thebibliography}

\end{document}